\documentclass[aps,prb,reprint,groupedaddress]{revtex4-2}
\usepackage{todo} 
\usepackage{siunitx} %
\usepackage{graphicx} 
\usepackage{dcolumn}  
\usepackage{amsmath}  
\usepackage{amssymb}  
\usepackage{physics}
\usepackage{color} 
\usepackage{xcolor}
\usepackage{hyperref} 
\usepackage{algorithmicx} 
\usepackage{xcolor}
\usepackage{soul}
\usepackage{comment}
\usepackage{xr}
\usepackage{orcidlink}

\begin{document}

\title{Tunable Emergent Gauge Fields from Skyrmions in a Quasicrystalline Lattice}

\author{Leandro M. Chinellato}
\email[]{lchinell@vols.utk.edu}
\affiliation{Department of Physics and Astronomy, The University of Tennessee, Knoxville, TN, 37996, USA}
\author{Flavia A.  G\'omez Albarrac\'in}
\email[]{albarrac@fisica.unlp.edu.ar}
\affiliation{Instituto de F\'isica de L\'iquidos y Sistemas Biol\'ogicos, CONICET, Facultad de Ciencias Exactas, Universidad Nacional de La Plata, 1900 La Plata, Argentina}
\affiliation{Departamento de Ciencias B\'asicas, Facultad de Ingenier\'ia, UNLP, La Plata, Argentina}
\author{Cristian D. Batista}
\affiliation{Department of Physics and Astronomy, The University of Tennessee, Knoxville, TN, 37996, USA}
\affiliation{Neutron Scattering Division and Shull-Wollan Center, Oak Ridge National Laboratory, Oak Ridge, TN, 37831, USA}
\author{Pablo S. Cornaglia~\orcidlink{0000-0003-4991-6573}}
\email[]{pablo.cornaglia@ib.edu.ar}
\affiliation{Centro Atómico Bariloche and Instituto Balseiro, CNEA, 8400 Bariloche, Argentina}
\affiliation{Consejo Nacional de Investigaciones Científicas y Técnicas (CONICET), Argentina}
\affiliation{Instituto de Nanociencia y Nanotecnología CNEA-CONICET}

\date{\today}

\begin{abstract}
We study magnetic skyrmions in a two-dimensional quasicrystalline lattice using a classical Heisenberg model with Dzyaloshinskii--Moriya interactions and an external magnetic field. The competition between the skyrmion--skyrmion repulsion and an emergent quasiperiodic pinning landscape gives rise to a sequence of distinct skyrmion lattice configurations as a function of field.
The resulting hierarchy of quasiperiodic pinning potentials, characterized by closely spaced quasi-degenerate minima, enables a quasi-continuous suppression of the skyrmion density as the saturation field is approached, in sharp contrast to the strongly first-order collapse of skyrmion crystals on periodic lattices. This provides a direct mechanism for controlling the topological charge and, consequently, the emergent gauge field for itinerant electrons. As a consequence, the Hall conductivity can be strongly modified with small changes in the magnetic field and driven smoothly to zero near saturation.
This field-controlled tunability, rooted in the underlying multistability, identifies quasicrystalline magnets as a platform for tunable topological textures, with potential applications in magnetic memory and magnetoelectronic response.
\end{abstract}

\maketitle

\section{Introduction}

Skyrmions are topologically protected magnetic textures characterized by a conserved topological charge. Owing to their robustness and compact size, they have been proposed as building blocks for magnetic memory devices. They have been observed in bulk materials, thin films, and multilayers, and are supported by a wide range of non-centrosymmetric systems, including B20 compounds, Heusler alloys, transition-metal thin films, and magnetic van der Waals (vdW) materials such as Fe$_3$GeTe$_2$~\cite{wu2022van,huang2022ferroelectric}. {More recently, skyrmions have also been identified in rare-earth magnets, both in the chiral compound EuPtSi~\cite{Khatua2025} and in centrosymmetric compounds such as Gd$_2$PdSi$_3$, Gd$_3$Ru$_4$Al$_{12}$~\cite{Kurumaji2019,Paddison2022,Hirschberger2019,Mo2025}, and EuAl$_4$~\cite{Arai2026}; in the latter, they emerge from competing exchange interactions and magnetic anisotropy in the absence of Dzyaloshinskii--Moriya interactions.}

The discovery of intrinsic magnetism in two-dimensional vdW crystals~\cite{lee2016ising,Burch2018} has established a versatile platform for exploring low-dimensional magnetism and spin-based functionalities. Although many vdW magnets order below room temperature, several compounds remain magnetically ordered above 300~K~\cite{Huang2020}. The magnetic ground states of these materials can be tuned through electrostatic gating, external fields, strain, pressure, or layer engineering. Because adjacent layers are only weakly bound, they can also be stacked with a controlled twist angle. The resulting Moiré superlattices enable precise control of the magnetic interactions, promote non-collinear spin textures~\cite{Chen2019,gibertini2019magnetic,Xu2022}, and can generate effective periodic potentials for skyrmions~\cite{Tong2018}.

The same stacking freedom can also produce quasiperiodic, rather than periodic, order. Graphene bilayers twisted by $30^\circ$, for instance, realize a dodecagonal quasicrystal hosting emergent Dirac physics~\cite{ahn2018dirac,yao2018quasicrystalline}. More generally, quasiperiodicity arises whenever mismatched lattice constants or twist angles prevent the formation of a commensurate supercell; in multilayers, independent twists can produce mutually incommensurate superlattices, realizing tunable Moiré quasicrystals~\cite{Hao2024}. Together with heterostructures combining materials of different symmetries and substrate-induced patterning, these routes offer realistic pathways to design quasiperiodic magnetic landscapes in vdW systems. In parallel, bulk compounds whose structure closely resembles that of a quasicrystal have been synthesized and shown to host rich magnetic ground states~\cite{Tamura2010,Kim2012,Woodland2026}.

Our aim is to identify how the quasicrystalline nature of the underlying lattice influences the arrangement and field evolution of skyrmion arrays. To this end, we use a minimal model in which skyrmions are stabilized by Dzyaloshinskii--Moriya interactions. The stabilization mechanism determines the skyrmion profiles and interactions, while quasiperiodicity supplies the hierarchy of local pinning environments central to our analysis. We therefore expect the geometric pinning mechanism identified here to operate more broadly, including in systems where skyrmions are stabilized by competing exchange or dipole--dipole interactions, provided that the textures remain sufficiently localized to resolve distinct local environments.

We first show that the quasiperiodic structure produces a position-dependent skyrmion energy landscape with a singular spectrum of closely spaced, quasi-degenerate minima. This hierarchy of pinning energies yields intrinsic multistability and allows the skyrmion density to be tuned by the magnetic field, offering a route to control the emergent gauge field of the texture and amplify its response to external fields. This mechanism arises from the interplay between the quasiperiodic pinning potential and the exponential localization of skyrmions, and has no analogue in periodic systems. We further examine how skyrmions order as the field is varied, distinguishing regimes dominated by inter-skyrmion repulsion from those in which quasiperiodic pinning governs the skyrmion arrangement. Finally, by coupling the spin texture to an itinerant electron bath, we show that the tunable skyrmion density translates, through the \emph{topological Hall effect} (THE), into sharp field control of the Hall conductivity.

The remainder of this paper is organized as follows. Section~\ref{sec:model} introduces the quasicrystalline lattice and the classical spin model with Dzyaloshinskii--Moriya interactions. Section~\ref{sec:polarized} shows that, owing to a residual Dzyaloshinskii--Moriya torque, the fully polarized state is replaced at high fields by a quasi-fully polarized state, and analyzes its gapped magnon spectrum, dynamical structure factor, and the extended or multifractal character of the magnon eigenstates. Section~\ref{sec:single} analyzes isolated skyrmion solutions and their energy spectrum across different pinning environments. In Sec.~\ref{sec:mc}, we present Monte Carlo simulations of the collective organization of skyrmions as a function of the external magnetic field. Section~\ref{sec:effmod} introduces an effective model for the low-energy skyrmion configurations near the critical field, built from the pinning landscape and skyrmion--skyrmion interactions. In Sec.~\ref{sec:THE}, we couple the spin texture to itinerant electrons and show how the field-tunable skyrmion density controls the topological Hall response~\cite{Mukherjee2023}. Finally, Sec.~\ref{sec:sumcon} summarizes our main findings and discusses potential implications for skyrmion-based devices.

\section{Model}\label{sec:model}
\begin{figure}
    \centering
    \includegraphics[width=0.48\textwidth]{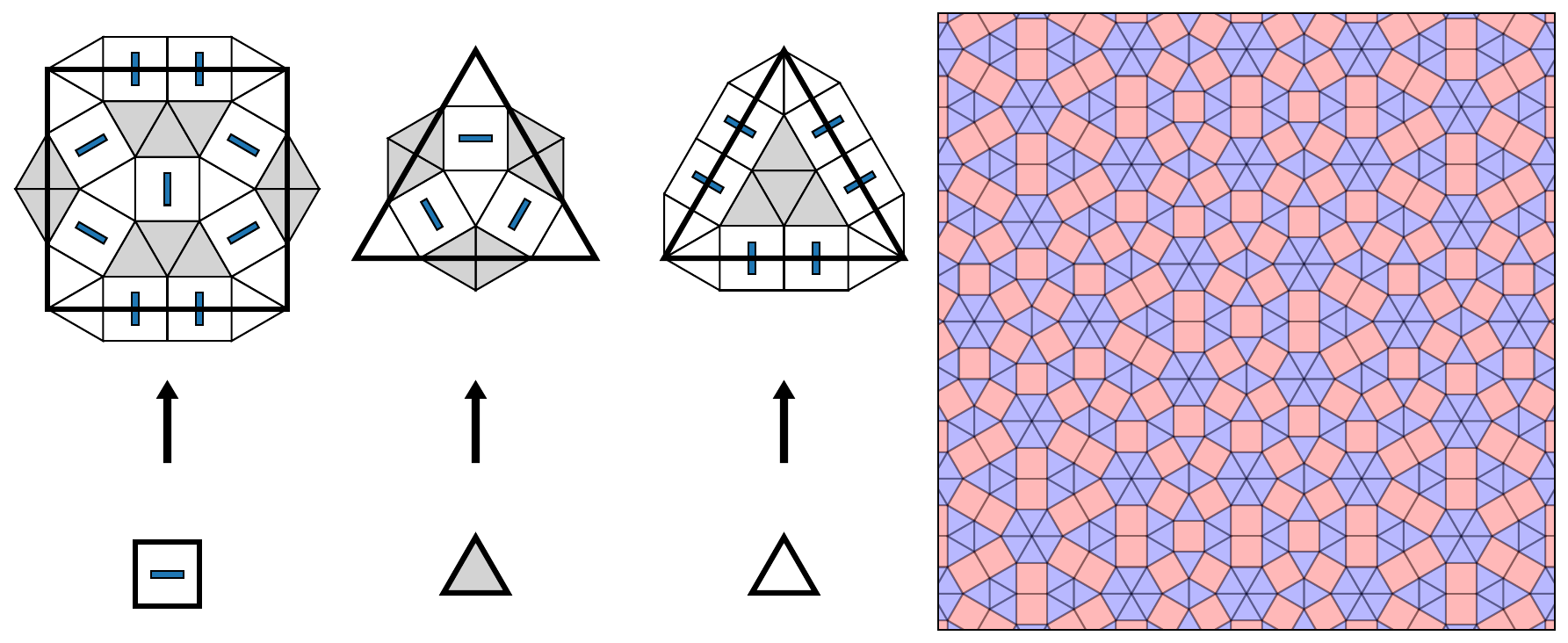}
    \caption{Inflation rules used to generate the quasicrystalline structure~\cite{barriotiling1998}.  The right panel shows a fragment of the quasicrystalline lattice.}   
    \label{fig:quasilattice}
\end{figure}

We consider a two-dimensional quasicrystalline lattice obtained by tiling the plane with equilateral triangles and squares~\cite{barriotiling1998}.
The structure is generated by the deterministic inflation rules shown in Fig.~\ref{fig:quasilattice}, applied to an initial square tile. Of the two tilings introduced in Ref.~\cite{barriotiling1998}, we use the quasiperiodic one, whose substitution matrix has a Pisot dominant eigenvalue.
In the thermodynamic limit, squares and triangles occupy equal total areas.
The construction is self-similar: starting from a single square, the sites where six triangles share a common vertex reproduce the full vertex set, rescaled by a factor of $2+\sqrt{3}$ and rotated by $90^\circ$ about the center of the square. 
In the following, we adopt the nearest-neighbor distance in the quasicrystal as the unit of length.

To construct the {spin} model we assign a classical magnetic moment $\mathbf{S}_i$ to each vertex of this quasicrystalline structure. The classical spin Hamiltonian reads
\begin{equation}\label{eq:mham}
    \mathcal{H} = -J \sum_{\langle i,j \rangle} \mathbf{S}_i \cdot \mathbf{S}_j + \sum_{\langle i,j \rangle}\mathbf{D}_{ij} \cdot (\mathbf{S}_i \times \mathbf{S}_j) - \sum_i \mathbf{H} \cdot \mathbf{S}_i,
\end{equation}
where the first term is the Heisenberg exchange interaction, the second term is the Dzyaloshinskii--Moriya (DM) interaction, and the third term is the Zeeman coupling with an external magnetic field $\mathbf{H} =H\hat{z}$. The sum $\langle i,j \rangle$ runs over nearest-neighbor pairs of sites. The Dzyaloshinskii--Moriya vectors $\mathbf{D}_{ij}$ are defined on the quasicrystalline lattice as
\begin{equation}
    \mathbf{D}_{ij} = D\left( \hat{\mathbf{u}}_{ij}\times \hat{\mathbf{z}} \right),
    \label{eq:DM}
\end{equation}
where $D$ is the strength of the DM interaction, {$\hat{\mathbf{u}}_{ij}= (\mathbf{r}_j - \mathbf{r}_i)/|\mathbf{r}_j - \mathbf{r}_i|$} is the unit vector pointing from site $i$ to site $j$, and $\hat{\mathbf{z}}$ is the unit vector pointing out of the plane. Notice that the Hamiltonian does not possess any continuous symmetry.

The DM interaction favors spiral spin arrangements and is typically a key ingredient for the stabilization of skyrmions. In the following we set $J$ as the energy unit, $D=1.08$, $|\mathbf{S}_i|=1$, and spin $S_z=1$ boundary conditions. The value $D = 1.08J$ is chosen to reduce the size of the skyrmion textures to a few lattice constants. The qualitative physics and topological properties are expected to be representative of the small-$D/J$ regime.

Successive inflations of the seed square generate square approximants to the quasicrystal that admit periodic tilings of the plane, enabling the use of periodic boundary conditions (PBC).
Each inflation step multiplies the linear size of the approximant by $2+\sqrt{3}$, and hence
its area by $(2+\sqrt{3})^2 \approx 13.9$. We consider two realizations. The first, which we call the {\it periodic approximant} is the level-3 inflation, a square of side $(2+\sqrt{3})^3 = 26+15\sqrt{3} \approx 51.98$ containing $2911$ sites, on which we impose PBC. 
The second, the {\it fixed spin patch}, is obtained by inflating to level 4 and cropping a $100\times100$ region (about $10,700$ points); since this patch cannot be tiled periodically, we instead apply fixed-spin boundary conditions, with the boundary spins aligned along the external magnetic field.

\section{Quasi-fully polarized state and magnon spectrum}\label{sec:polarized}

We are first interested in studying the high-field regime, where the spins are nearly aligned with the applied field $\mathbf{H} = H\hat{\mathbf{z}}$. A natural starting point would be the \emph{fully polarized} (FP) configuration $\mathbf{S}_i = S\hat{\mathbf{z}}$ for all $i$. On a regular lattice with sufficient point-group symmetry, this state is an exact stationary point of $\mathcal{H}$ above the saturation field, and linear spin-wave theory about it yields a well-defined magnon spectrum. On the quasicrystal this is no longer the case: The fully polarized state is \emph{never} an exact ground state of $\mathcal{H}$ at finite $H$, neither classically nor quantum-mechanically.

The N\'eel-type DM vector $\mathbf{D}_{ij}=D(\hat{\mathbf{u}}_{ij}\times\hat{\mathbf{z}})$ lies in the plane, and the variation of the {DM}  energy with respect to a small in-plane tilt $\delta\mathbf{S}_i^{\perp}$ of site $i$ about the polarized configuration is
\begin{equation}
\left.\frac{\partial \mathcal{H}_{\rm DM}}{\partial \mathbf{S}_i^{\perp}}\right|_{\rm FP}
= {S\,D\,} \sum_{j\in\partial i}\hat{\mathbf{u}}_{ij},
\label{eq:torque}
\end{equation} 
where $\partial i$ denotes the set of nearest neighbors of site $i$. On crystalline lattices with inversion or sufficient rotational symmetry at every site, $\sum_{j\in\partial i}\hat{\mathbf{u}}_{ij}=0$ and the torque Eq.~\eqref{eq:torque} vanishes identically, restoring the polarized state as a stationary point above $H_{c}$. Although the quasicrystalline lattice lacks the symmetry required to enforce
$
\sum_{j\in\partial i}\hat{\mathbf{u}}_{ij}=0,
$
the local environments still retain a high degree of geometric balance. The vectors $\hat{\mathbf{u}}_{ij}$ are distributed in an almost symmetric fashion around each site, so their sum does not vanish identically, but it is strongly suppressed by a near-cancellation among neighboring bonds. As a result, the {DM} torque in Eq.~\eqref{eq:torque} is generically finite but parametrically small.
Physically, this implies that the fully polarized state is no longer an exact stationary point, yet it remains an excellent zeroth-order approximation. The residual in-plane torque induces a weak canting of the spins, whose magnitude is controlled by the small imbalance in $\sum_{j}\hat{\mathbf{u}}_{ij}$. This canting is  suppressed by the dominant longitudinal field and exchange contributions, which favor alignment along $\hat{z}$.
Consequently, the system realizes a quasi-fully polarized state (QFP).

\begin{figure}[ht]
\centering
\includegraphics[width=\columnwidth]{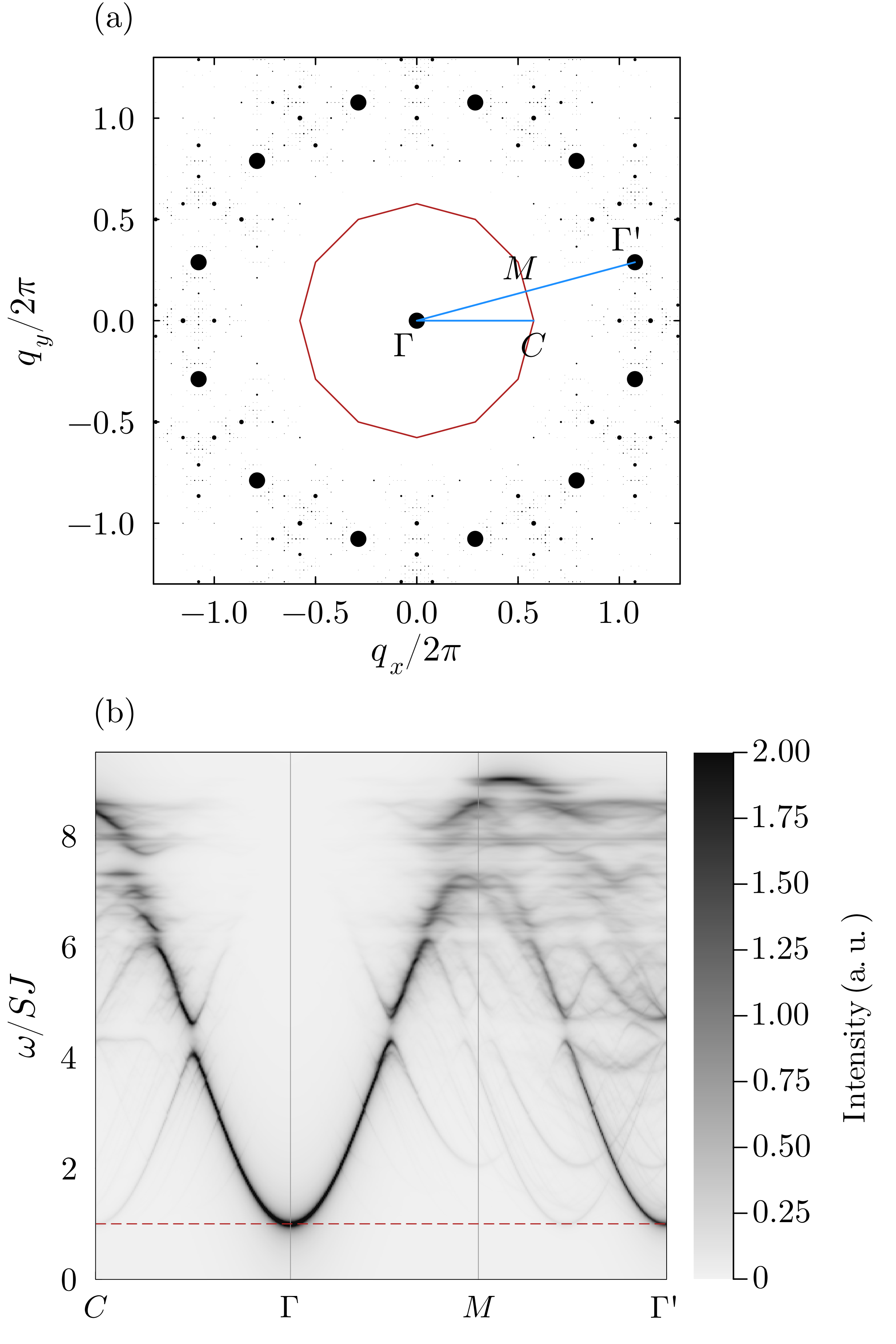}
\caption{(a) Dodecagonal pseudo-Brillouin zone bounded by the bisectors (red) of the twelve first-order Bragg peaks $\mathbf{g}^{I}$ (dots),  with the path $C$--$\Gamma$--$M$--$\Gamma'$ (blue). The momenta $(q_x,q_y)$ are in units of $2\pi$.
(b) One-magnon dynamical structure factor $\mathcal{S}^{\perp}(\mathbf{q},\omega)$ in the QFP regime, within the fully polarized spin-wave approximation. Obtained at $H=0.9957J$  by unfolding~\cite{Zhang2022} the bands of the $N=2911$-site approximant under periodic boundary conditions. The reciprocal-space path $C$--$\Gamma$--$M$--$\Gamma'$ follows the pseudo-Brillouin zone defined by the strongest Bragg peaks $\mathbf{g}^{I}$ of the lattice, with $C$ a corner of the PBZ, $M=\mathbf{g}^{I}/2$ and $\Gamma'=\mathbf{g}^{I}$. The dashed line marks $\omega=H$, and the Lorentzian broadening is $\eta=0.03JS$. }
\label{fig:dssf_fullypolarized}
\end{figure}

As presented in Section~\ref{sec:effmod}, we determine this state numerically by minimizing the energy $\mathcal{H}$ for each applied magnetic field $H$, finding that the critical field $H_c\approx 0.9957$ is the field above which the minimum belongs to the zero-skyrmion sector. For $H\gtrsim H_c$ the deviations from full polarization are everywhere small, $|\delta\mathbf{S}_i^\perp|/S\ll 1$.
\begin{figure*}[ht]
    \centering
\includegraphics[width=0.82\linewidth]{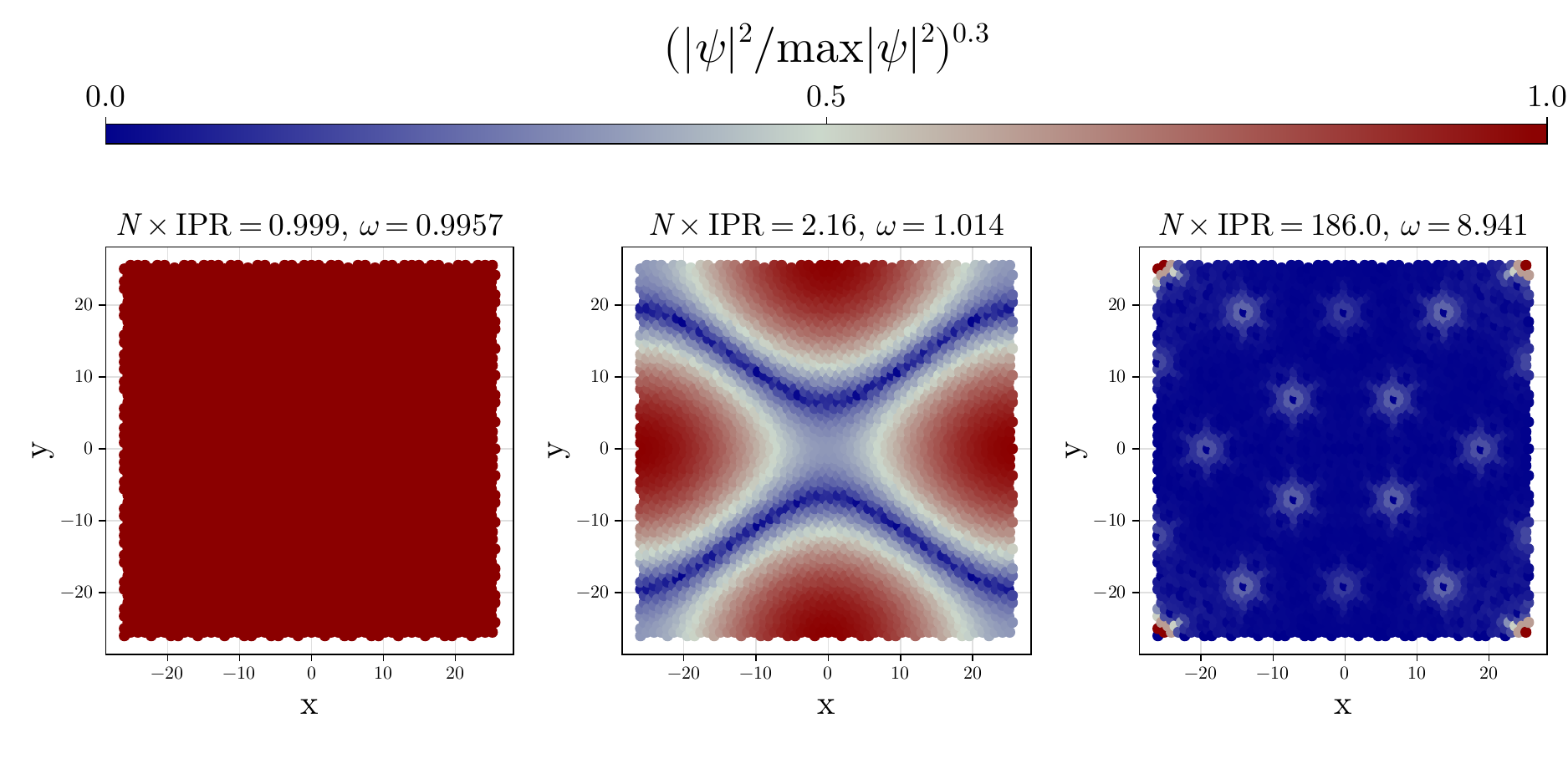}
    \caption{Squared probability amplitude $|\langle i|\psi_n\rangle|^2$ for representative magnon eigenstates of the quasicrystal approximant at the critical magnetic field $H=H_c$. The {amplitudes are} displayed {using} the power-law color scale $(|\psi|^2/\max |\psi|^2)^{0.3}$ to enhance the visibility of {low-intensity features}. Each panel is labeled by the eigenstate energy $\omega$ and the scaled participation ratio $N\times\mathrm{IPR}$.}
    \label{fig:square_amp}
\end{figure*}

The same mechanism appears naturally in the semiclassical $1/S$ expansion. A Holstein--Primakoff (HP) expansion of the DM term about the fully polarized state generates terms linear in the bosonic HP operators $a_i$ and $a_i^\dagger$, with coefficients proportional to $D\sum_{j\in\partial i}\hat{\mathbf{u}}_{ij}$ (the bosonic counterpart of the classical torque). These linear terms signal that the $\hat{\mathbf{z}}$-polarized state is not the true vacuum of the spin-wave theory. Above $H_c$ they are accordingly neglected, and the resulting quadratic bosonic Hamiltonian takes the simple tight-binding form,
\begin{equation}
\mathcal{H}_{\rm sw}
= J S\sum_{\langle ij\rangle}
\bigl(a_i^\dagger a_i + a_j^\dagger a_j
- \, a_i^\dagger a_j - \, a_j^\dagger a_i\bigr)
+ H \sum_i a_i^\dagger a_i.
\label{eq:TB}
\end{equation}
Since the DM vectors defined in Eq.~\eqref{eq:DM} are perpendicular to $\hat{\mathbf{z}}$, i.e.\ $D_{ij}^z=0$, no anomalous (Bogoliubov) terms are generated about the $\hat{\mathbf{z}}$-polarized state, and $\mathcal{H}_{\rm sw}$ reduces directly to a tight-binding Hamiltonian shifted by the Zeeman field.
Under this approximation, Eq.~\eqref{eq:TB} defines a single-particle tight-binding problem on the periodic approximant, whose $N=2911$ magnon bands are folded into the Brillouin zone of the square supercell, of linear size $2\pi/L\simeq0.12$, where $L=(2+\sqrt{3})^3$ is the side of the approximant. The folded bands carry no information on the momentum transfer $\mathbf{q}$ probed in a scattering experiment, which we recover by band unfolding following Ref.~\cite{Zhang2022}. The one-magnon dynamical structure factor per site, $\mathcal{S}^{\perp}=\mathcal{S}^{xx}+\mathcal{S}^{yy}$, reads
\begin{align}
\mathcal{S}^{\perp}(\mathbf{q},\omega)
&=\frac{1}{N}\sum_{ij}e^{-i\mathbf{q}\cdot(\mathbf{r}_i-\mathbf{r}_j)}\!\int\!\frac{dt}{2\pi}\,e^{i\omega t}\,\langle\mathbf{S}^{\perp}_i(t)\cdot\mathbf{S}^{\perp}_j(0)\rangle \nonumber\\
&\simeq-\frac{S}{\pi}\,\mathrm{Im}\,\langle\mathbf{q}|\big(\omega+i0^{+}-\mathcal{H}_{\rm sw}\big)^{-1}|\mathbf{q}\rangle,
\label{eq:dssf_unfolded}
\end{align}
where the second line holds at $T=0$ to leading order in $1/S$ ($S_i^{+}\simeq\sqrt{2S}\,a_i$), and $|\mathbf{q}\rangle=N^{-1/2}\sum_i e^{i\mathbf{q}\cdot\mathbf{r}_i}|i\rangle$ is a plane wave on the lattice sites. Since $\mathcal{H}_{\rm sw}$ commutes with supercell translations, $|\mathbf{q}\rangle$ couples only to the Bloch sector $\mathbf{K}=\mathbf{q}-\mathbf{G}$, with $\mathbf{G}$ a supercell reciprocal-lattice vector, and each folded band $\omega_n(\mathbf{K})$ contributes at $\mathbf{q}$ with the weight of its plane-wave component [see the Supplemental Material (SM) Sec.~I~\cite{SM}].

The resulting dynamical structure factor, shown in Fig.~\ref{fig:dssf_fullypolarized}(b), exhibits a fully gapped magnon spectrum, with a spectral gap  $\Delta = H$. Wave vectors are referred to the pseudo-Brillouin zone (PBZ) [see Fig.~\ref{fig:dssf_fullypolarized}(a)] bounded by the bisectors of the twelve strongest Bragg peaks $\mathbf{g}^{I}$ of the lattice structure factor $I(\mathbf{q})$~\cite{Zhang2022}. The spectral weight is concentrated in a sharp, dispersive branch that starts at $\omega=H$ at $\Gamma\equiv(0,0)$, where the uniform mode is an exact eigenstate, with $\omega\simeq H+JS\bar{z}|\mathbf{q}|^2/4$ and average coordination $\bar{z}=5.07$. The branch also shows an avoided crossing at $4.1JS\lesssim\omega\lesssim4.7JS$.

To fully characterize the spatial structure of the magnon eigenstates, we compute the inverse participation ratio (IPR),
\begin{equation}
\mathrm{IPR}_n = \sum_i |\psi_n(i)|^4.
\end{equation}
{For extended states, the inverse participation ratio scales as $\mathrm{IPR}\sim 1/N$, while for localized states it remains finite, $\mathrm{IPR}\sim\mathcal{O}(1)$.} We find that the lowest-energy magnon modes follow $\mathrm{IPR} \sim 1/N$, indicating that they are spatially extended. In contrast, higher-energy states exhibit multifractal scaling, $\mathrm{IPR} \sim N^{-\alpha}$ with $0<\alpha<1$, consistent with the phenomenology of quasicrystalline systems~\cite{Suck2013}.

The spatial character of the magnon eigenstates is illustrated in Fig.~\ref{fig:square_amp}, which shows the squared probability amplitude $|\langle i|\psi_n\rangle|^2$ for representative examples. Specifically, we compare the lowest-energy, most spatially extended eigenstate with the most localized one (see the SM~\cite{SM} for additional examples).

The momentum-resolved spectrum of Fig.~\ref{fig:dssf_fullypolarized}(b) is consistent with this real-space picture. The extended low-energy states carry a well-defined plane-wave component and form the sharp branch around $\Gamma$. At higher energies, the weight of a given $\mathbf{q}$ is shared among an increasing number of multifractal eigenstates, which fragments the branch above $\omega\approx6JS$ and leaves only faint, weakly dispersive features above $\omega\approx7.5JS$.

\section{Single-skyrmion solutions}\label{sec:single}

\begin{figure}[t]
\centering
\includegraphics[width=0.45\textwidth]{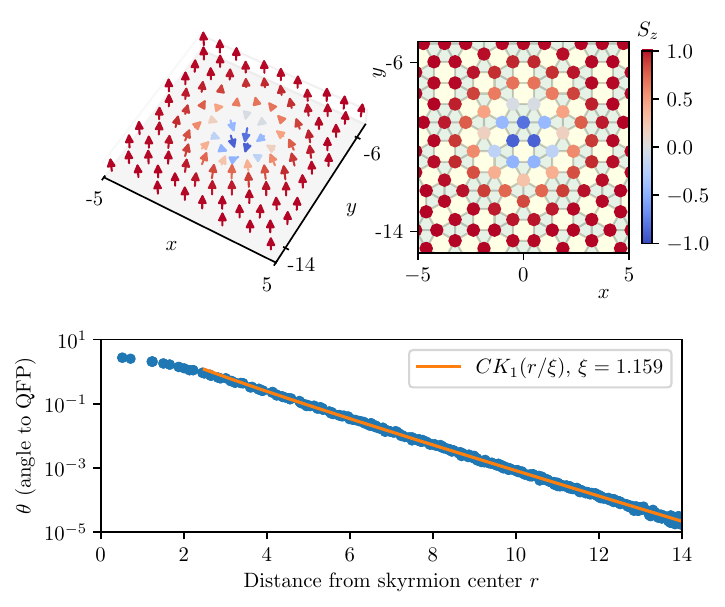}
\caption{Single skyrmion on the quasicrystalline lattice at $H=0.9894$.
Upper left: spin texture, with arrows indicating the magnetic moments and colors the out-of-plane component $S_z$. 
Upper right: $S_z$ on the lattice sites, with the square--triangle tiling in the background. 
Lower: angle $\theta$ between each spin and the corresponding spin of the quasi-fully polarized (QFP) state at the same field, as a function of the distance $r$ from the skyrmion center. 
The solid line is $C\,K_1(r/\xi)$, with $\xi(H)$ fixed by the magnon dispersion (see SM~\cite{SM}) and
$C=10.2$ the only fitted parameter ($2.5<r<14$).}
\label{fig:single_skyrmion}
\end{figure}

In addition to the quasi-fully polarized ground state, the system supports
metastable single-skyrmion solutions on the QFP state at fields $H \approx H_c$. Figure~\ref{fig:single_skyrmion} shows a representative single-skyrmion solution obtained by minimizing the spin Hamiltonian with the Broyden-Fletcher-Goldfarb-Shanno (BFGS) algorithm~\footnote{Each magnetic moment is represented as a unit-length classical vector parametrized using a skew-symmetric matrix~\protect\cite{ivanov2021fast}.}.
The system energy is expressed in terms of the matrix parameters and its gradient is computed analytically. Energy minimization is performed using the BFGS algorithm as implemented in the Ensmallen library~\cite{curtin2021ensmallen}, which uses Armadillo for linear algebra operations~\cite{sanderson2016armadillo}.  See
Ref.~\cite{cornaglia2023unveiling} for additional details. 

The minimization starts from a QFP configuration with a small cluster (typically three) of neighboring spins inverted relative to the background. Since the BFGS algorithm relies solely on local gradient and curvature information, it converges to the nearest local energy minimum, thereby preserving the topological sector of the initial configuration.

As shown in Fig.~\ref{fig:single_skyrmion}, the resulting configuration corresponds to a single skyrmion embedded in a QFP background. 
The upper left panel displays the magnetic moment texture, with an inverted core whose orientation varies with the radial distance. 
Although the overall profile resembles a conventional skyrmion, small deviations reflect the local
environment of the quasicrystalline lattice. The upper right panel shows the spatial distribution of~$S_z$, revealing a localized structure with $S_z \approx -1$ at the center and $S_z \approx +1$ far from it.
The tail of the skyrmion profile can be understood from the gapped magnon spectrum discussed in Section~\ref{sec:polarized}. 
Far from its core, the skyrmion is a weak perturbation of the QFP state, and its spin deviation is
the static linear response of that state, i.e., the zero-frequency magnon propagator.

The perturbation generated by a skyrmion on the QFP background can be characterized by the angle $\theta_i$ between each spin and the corresponding spin of the QFP state at the same field, as a function of the distance~$r$ from the
skyrmion center~\footnote{The skyrmion position is computed as the centroid weighted by $1-S_z$.} (see lower panel of Fig.~\ref{fig:single_skyrmion}). 
Outside the core, $\theta$ follows $C\,K_1(r/\xi)$, where $K_1$ is the modified Bessel function of the second kind.
The decay length $\xi\simeq \sqrt{JS\bar{z}/4H} $ is set by the gap and stiffness of the magnon spectrum (see SM~\cite{SM}), and the amplitude $C$ is the only fitted parameter.
In the language of virtual magnons, two skyrmions interact by exchanging these gapped excitations, which gives a repulsive
potential $V(r) = V_0K_0(r/\xi)$ with the same length $\xi$ (Sec.~\ref{sec:effmod}). 
The gap thus explains why skyrmions behave as nearly independent particles near $H_c$: instead of forming a regular
lattice set by their mutual repulsion, they occupy the pinning sites selected by the quasicrystalline tiling.

The skyrmions carry a well-defined topological charge. To compute it, we map each three-dimensional unit vector $\mathbf{S}_i/|\mathbf{S}_i| = (n_{x,i}, n_{y,i}, n_{z,i})$, which specifies the direction of the magnetic moment at site $i$, onto a coherent state in the fundamental irreducible representation of SU(2), exploiting the standard isomorphism $S^2 \simeq \mathbb{CP}^1$. The mapping is explicitly realized by the spinor
\begin{equation}
|\Psi_i \rangle =
\begin{bmatrix}
\cos\left(\frac{\vartheta_i}{2}\right) e^{-i \frac{\varphi_i}{2}} \\
\sin\left(\frac{\vartheta_i}{2}\right) e^{i \frac{\varphi_i}{2}}
\end{bmatrix},
\end{equation}
where $\vartheta_i = \arccos(n_{z,i})$ and $\varphi_i = \arctan(n_{y,i}, n_{x,i})$.

The solid angle associated with each plaquette can be obtained from the Berry phase accumulated along a closed loop of spinors surrounding that plaquette (triangular or square tile). The Berry phase is given by
\begin{equation}
\Omega_p = 2\arg\left( \prod_{j=1}^{N_p} \langle \Psi_j | \Psi_{j+1} \rangle \right),
\end{equation}
where the index $j$ runs over the $N_p$ spinors around the plaquette in a clockwise manner, and $\langle \Psi_j | \Psi_{j+1} \rangle$ denotes the overlap between consecutive spinors (with $N_p + 1 \equiv 1$ for closure).

Finally, the total topological charge of the system is obtained by summing the Berry phases over the elementary faces of the quasicrystalline tiling, 
namely the minimal triangular and square plaquettes~\cite{BERG1981412}
\begin{equation}
Q = \frac{1}{4\pi} \sum_{p} \Omega_p.
\end{equation}
For periodic boundary conditions, or with the boundary spins fixed parallel to the applied magnetic field, the topological charge $Q$ is quantized to integer values. In particular, the single-skyrmion solution has $Q= \pm 1$, confirming its topological character.

\begin{figure}[t]
    \centering
    \includegraphics[width=0.45\textwidth]{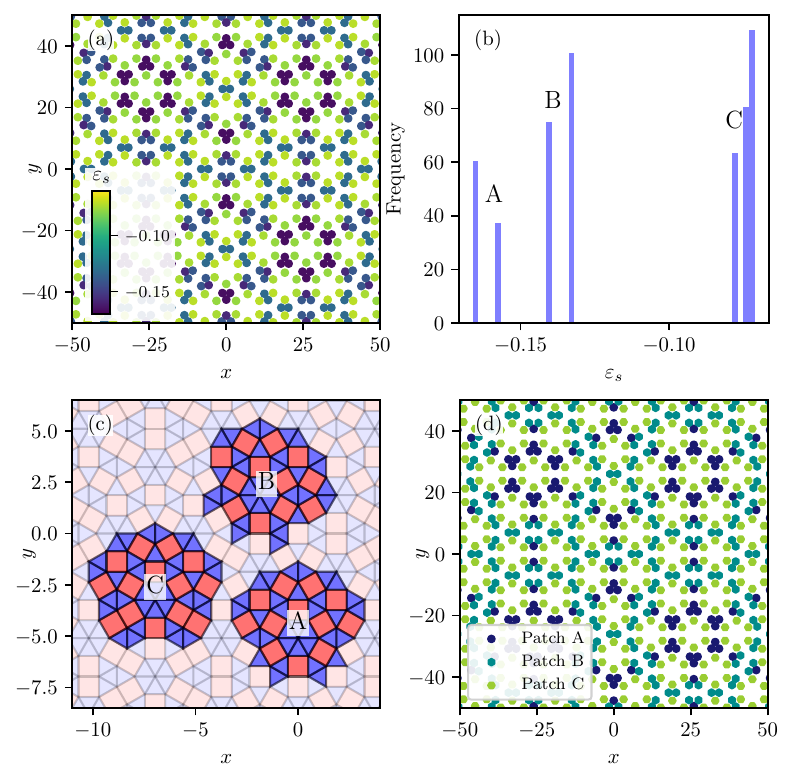}
    \caption{(a) Single skyrmion energy as a function of the position for the fixed-spin boundary conditions patch {at  magnetic field $H=0.9894$ }. (b) {Histogram of single skyrmion energy}. The three  {labeled peaks correspond to skyrmions located at} the center of the patches {shown} in (c). (d) Locations of the $A$, $B$, and $C$ patches, {including their rotationally equivalent realizations within the quasicrystal.} }
    \label{fig:pinning}
\end{figure}
Skyrmions stabilize at specific local environments within the quasicrystal. To characterize these environments, we define a {\it patch} as a finite region centered at the skyrmion core which includes the surrounding lattice structure up to a given radius. Because the quasicrystal was constructed through a primitive inflation rule, these local configurations recur throughout the structure in a quasiperiodic manner~\cite{baake2013aperiodic}.

Figure \ref{fig:pinning}(a) presents the metastable locations of single skyrmions, where the color code indicates the single skyrmion energy $\varepsilon_s$. This energy is defined as the difference between the energy of the single skyrmion state and that of the quasi-fully polarized state: $\varepsilon_s=E_{1s}-E_{QFP}$, where $E_{1s}$ is the energy of a single skyrmion embedded in quasi-fully polarized background, and $E_{QFP}$ is the energy of the quasi-fully polarized configuration.

A clear pattern emerges, with three distinct color groupings corresponding to three energy sectors. This separation is evident in the single-skyrmion energy histogram of Fig.~\ref{fig:pinning}(b), which displays three clusters of peaks labeled $A$, $B$, and $C$. Each cluster is associated with a specific local environment in the quasicrystal where skyrmions preferentially stabilize. The energies labeled $A$, $B$, and $C$ correspond to distinct local patches, also labeled $A$, $B$, and $C$ in Fig.~\ref{fig:pinning}(d). Figure~\ref{fig:pinning}(c) shows the spatial distribution of these patches throughout the quasicrystal, including their rotated equivalents.

 The substructure observed in the peaks of the histogram in Fig.~\ref{fig:pinning}(b) originates from the distinct extensions of each local patch. Although a given patch appears infinitely many times throughout the quasicrystal, no two extended environments are strictly identical, leading to a hierarchical organization of local environments surrounding the skyrmions as the patch radius increases. Because skyrmions are exponentially localized objects, different extensions of the same patch produce energy differences that decrease exponentially with the size of the extension, and the single-skyrmion energy becomes a local quantity, set entirely by the lattice environment within a few decay lengths $\xi$ of the core. The energy spectrum is therefore expected to be singular, reflecting both this exponential localization and the dense, quasiperiodic variation of local environments in the underlying quasicrystal. Similar behavior has been reported for chiral solitons in Fibonacci quasicrystals~\cite{cornaglia2024quasicrystalline}.

The locality of the single-skyrmion energy further implies that the pinning energies reported here are independent of the choice of system and boundary conditions, provided the skyrmion core sits at a distance $\gg\xi$ from the boundary, which is the case for all interior positions considered. As we will see in the next section, this hierarchy of energy scales gives rise to a dense set of nearly degenerate configurations, thereby enabling a quasi-continuous evolution of the skyrmion density below the critical field $H_c$.

\section{Multiskyrmion Configurations}\label{sec:mc}

To obtain low-energy skyrmion configurations, we carried out Monte Carlo simulations combining Metropolis updates with microcanonical (overrelaxation) moves~\cite{creutz1987overrelaxation}. The simulations were performed on both realizations of the quasicrystal: the fixed-spin patch ($\sim 10{,}700$ sites) and the periodic approximant ($N=2911$, with periodic boundary conditions).
The temperature was gradually reduced from $T/J=3$ to $T/J=6.5\times10^{-4}$ in 80 steps using a geometric simulated annealing schedule with \(T_{i+1}=0.9T_i\). For each value of the external magnetic field, we performed 10--20 independent simulations with different random seeds. Each simulation consisted of $10^5$ Monte Carlo steps (MCS) for thermalization, followed by $2\times10^5$ MCS for measurements and sampling of spin configurations.

\begin{figure}
    \centering
    \includegraphics[width=0.45\textwidth]{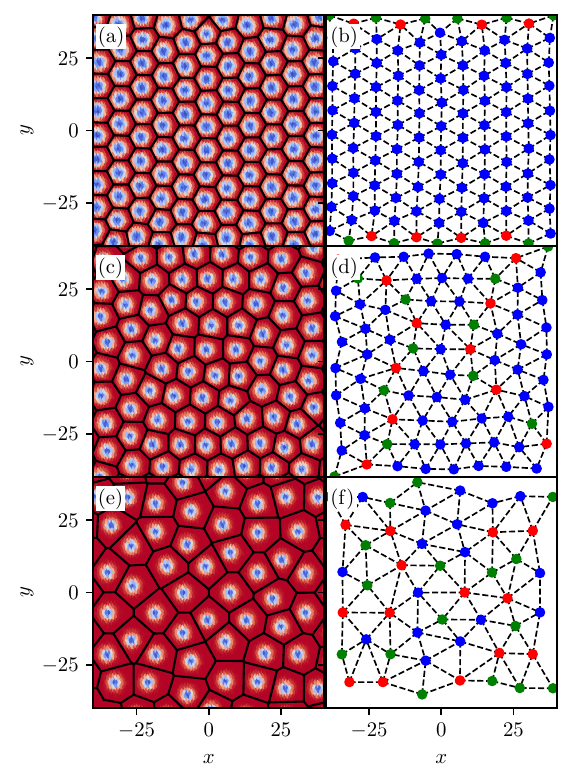}
   \caption{Low-temperature Monte Carlo configurations {for} the fixed-spin patch at (a,b) $H=0.6$,  (c,d) $H=0.9$, and (e,f) $H=0.994$ . Left panels show the spin texture, {with colors representing $S_z$, and the corresponding Voronoi tessellation overlaid. Right panels show the corresponding skyrmion network, with sites colored according to their} coordination number.}
    \label{fig:configsMC}
\end{figure}

Three representative examples of low-temperature skyrmion configurations on the fixed-spin patch are presented in Fig. \ref{fig:configsMC} for different values of the magnetic field.
At low magnetic fields, the skyrmion density is limited by the repulsive skyrmion-skyrmion interaction, which dominates over the pinning potential. In this regime, the skyrmions {form a} triangular lattice, as shown in Fig. \ref{fig:configsMC}(a). A Voronoi analysis reveals coordination defects near the sample boundaries [see Fig. \ref{fig:configsMC}(b)], which arise from the fixed-spin boundary conditions of the patch.

As the magnetic field increases, the skyrmion density decreases and the pinning potential becomes increasingly dominant relative to the skyrmion-skyrmion interaction. As a consequence, the positional disorder of the skyrmion lattice increases and the density of coordination defects increases, as illustrated in Fig. \ref{fig:configsMC}(c), where domain boundaries separate triangular domains with different crystallographic orientations.

\begin{figure}
    \centering
    \includegraphics[width=0.45\textwidth]{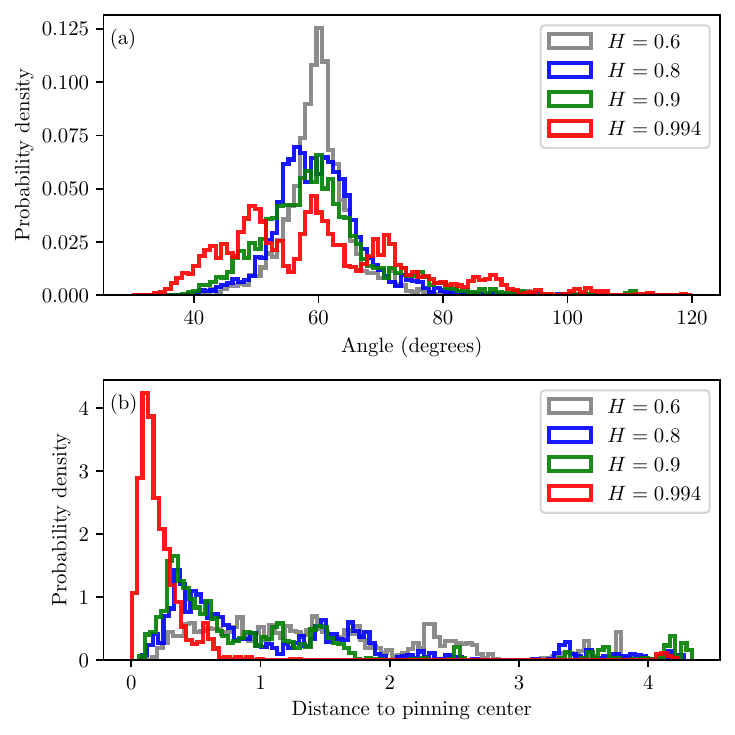}
    \caption{a) 
    Probability distribution of the angles between the vectors connecting each skyrmion to two neighboring skyrmions. b) Distribution of the minimum distance between each skyrmion and the nearest pinning center.
    Results are shown for different external magnetic fields and are obtained from the lowest temperature Monte Carlo configurations averaged over 10 independent simulated annealing runs for each field.}
    \label{fig:distrib}
\end{figure}

For higher magnetic fields near $H_c$ the skyrmion density decreases  further while coordination defects proliferate, leading to a skyrmion configuration reminiscent of a liquid. 
Despite their liquid-like appearance, these configurations exhibit two distinctive features.
First, short-range square-lattice correlations emerge. This trend is evident from the distribution of nearest-neighbor bond angles shown in Fig. \ref{fig:distrib}.
At low magnetic field ($H=0.6$), the distribution exhibits a pronounced peak at $60^{\circ}$, as expected for an ideal triangular lattice. As the field increases, the distribution broadens, and for $H\sim 0.9$, secondary peaks emerge near $51.4^\circ$ and $72^\circ$, consistent with the presence of sevenfold and fivefold coordination defects, respectively. Close to the critical field, additional peaks appear near $45^\circ$ and $90^\circ$, indicating the emergence of local square-lattice order.
Second, skyrmions become increasingly localized at pinning centers as the magnetic field approaches $H_c$. The distribution of the minimum distance  between skyrmions and pinning centers, shown in Fig. \ref{fig:distrib}(b), indicates that most skyrmions lie within one lattice spacing of a pinning center as $H \to H_c$.

For magnetic fields $H\approx H_c$ the simulated annealing yields low-temperature states in which all skyrmions are localized at pinning centers. However, the competition between skyrmion-skyrmion interactions and the complex pinning energy landscape gives rise to rugged energy landscape with many local minima. As a result, the equilibrium skyrmion number cannot be reliably inferred from a single annealing run. Once formed, skyrmions are protected by a finite topological energy barrier and are therefore difficult to annihilate. Consequently, identical annealing protocols initiated from different random seeds may converge to metastable states containing different numbers of skyrmions.

To overcome the metastability and sample-to-sample fluctuations in skyrmion number near $H_c$, we introduce in the next section a complementary approach that enables a controlled exploration of low-energy configurations.

\section{Skyrmion configurations for $H\lesssim H_c$}\label{sec:effmod}

In this section, we investigate the equilibrium configurations of skyrmions in the regime where the external magnetic field $H$ lies slightly below the critical field $H_c$. We examine the interplay between the local pinning potential and the skyrmion-skyrmion interaction to identify the energetically favorable configurations. Assuming that skyrmions are localized at pinning centers, we construct an effective energy model for point-like skyrmions occupying discrete pinning sites and interacting through the effective skyrmion--skyrmion interaction derived in the previous section. Simulated annealing is then employed to identify the low-energy skyrmion configurations as a function of the magnetic field.

\begin{figure}[t]
    \centering
    \includegraphics[width=0.5\textwidth]{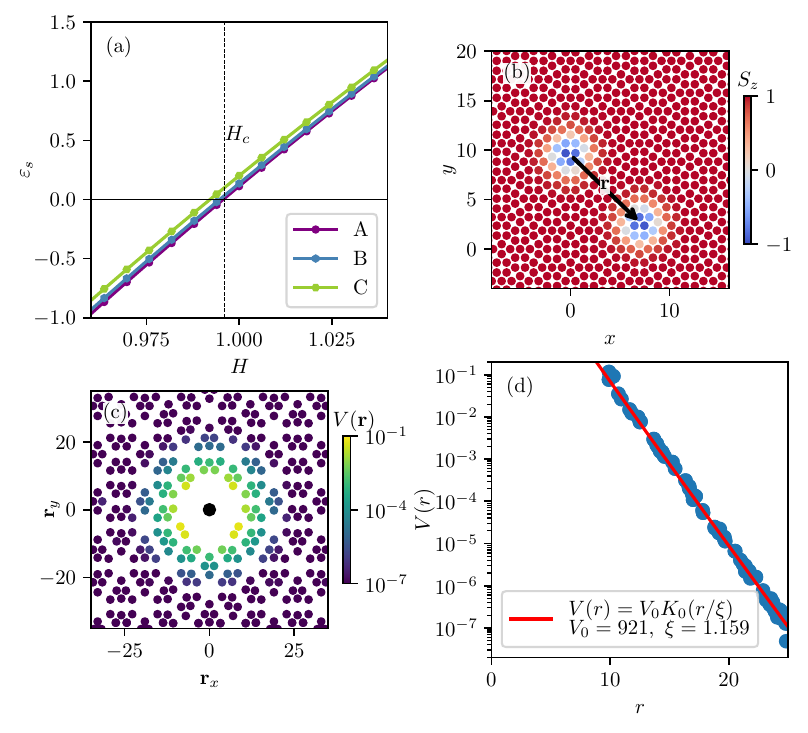}
    \caption{(a) Single-skyrmion energy $\varepsilon_s$ as a function of the external magnetic field $H$ for three inequivalent pinning sites A, B, and C of the quasicrystalline lattice (see Fig.~\ref{fig:pinning}). 
    The dashed line marks the critical field $H_c = 0.9957$, above which isolated skyrmions are energetically unfavorable.
    (b) Magnetic configuration of two skyrmions embedded in the QFP background. The color scale indicates $S_z$, and $\mathbf r$ is the separation vector between the skyrmion centers.
(c) Interaction energy $V(\mathbf r)$ for a skyrmion fixed at $\mathbf R_0 = (0, 9.4)$ (black disk) and a second skyrmion placed at the relative positions $\mathbf r$ indicated by the small disks, at $H = 0.9894$.
(d) $V$ as a function of the separation $|\mathbf r|$, extracted from panel (c). The red line is the linear-response form $V_0K_0(|\mathbf r|/\xi)$, with $\xi = 1.159$ fixed by the magnon dispersion as in Fig.~\ref{fig:single_skyrmion} and $V_0 = 921$ the only fitted parameter. The scatter around the line reflects the dependence of $V$ on the direction
of $\mathbf r$ and on the local environment.}
    \label{fig:interactions}
\end{figure}

Figure~\ref{fig:interactions}(a) presents the single-skyrmion energy as a function of the external magnetic field for skyrmions centered at three distinct positions in the quasicrystalline lattice, corresponding to patches $A$, $B$, and $C$ defined in Sec.~\ref{sec:single}. For each location, the single-skyrmion energy decreases as the magnetic field is reduced and eventually becomes negative. The skyrmion centered on an $A$ patch has the lowest energy and therefore becomes stable with respect to the QFP state at the highest magnetic field. The critical field $H_c$ is defined as the largest magnetic field at which a single skyrmion has lower energy than the topologically trivial QFP state.

The interaction potential at fields $H \simeq H_c$ can be estimated by placing two skyrmions at distinct pinning locations within an otherwise QFP background and relaxing the spin configuration using the BFGS method. For sufficiently large separations, the BFGS relaxation preserves the skyrmions near their initial pinning centers, allowing the interaction energy to be extracted reliably. The interaction potential is then defined as the nonadditive contribution to the total energy,
\begin{equation}
V(\mathbf{r}) = \varepsilon_{2s}(\mathbf{R},\mathbf{R}^\prime) - \big[\varepsilon_{s}(\mathbf{R}) + \varepsilon_{s}(\mathbf{R}^\prime)\big],
\end{equation}
where $\mathbf{R}$ and $\mathbf{R}^\prime$ denote the skyrmion positions, $\mathbf{r}=\mathbf{R}-\mathbf{R}^\prime$ is their relative separation, and $\varepsilon_s(\mathbf{R}) = E_{1s}(\mathbf{R}) - E_{QFP}$ and $\varepsilon_{2s}(\mathbf{R},\mathbf{R}^\prime) = E_{2s}(\mathbf{R},\mathbf{R}^\prime) - E_{QFP}$ are the single- and two-skyrmion energies, measured relative to the QFP state [Fig.~\ref{fig:interactions}(b)].

Figure~\ref{fig:interactions}(c) shows the interaction potential as a function of the relative position $\mathbf{r}$. The potential is determined at the discrete relative {positions} {at which} pairs of skyrmions {can} be stabilized. {The absence of data} near $(0,0)$ {reflects the fact that} the interaction {is} too strong at these short distances for the skyrmions to remain pinned at their original positions. As expected, the interaction decreases exponentially with increasing {inter-skyrmion} distance.

To find {approximate ground states} for fields near $H_c$, we {describe} the system as {point-like skyrmions confined to the} metastable pinning positions $\{\mathbf{R}_i\}$. Let $n_i\in\{0,1\}$ indicate the presence ($n_i=1$) or absence ($n_i=0$) of a skyrmion at site $i$. The total energy of a configuration is approximated as
\begin{equation}
E_{\mathrm{tot}} = \sum_i n_i\, \varepsilon_s(\mathbf{R}_i; H) + \frac{1}{2} \sum_{i \neq j} n_i n_j\, V\left( |\mathbf{R}_i - \mathbf{R}_j| \right),
\end{equation}
where the field-dependent single-skyrmion energy is modeled as
\begin{equation}
\varepsilon_s(\mathbf{R}_i; H) = \varepsilon_s(\mathbf{R}_i;H_c) + \alpha (H - H_c),
\end{equation}
with $\alpha>0$, where the magnetic field plays the role of a chemical potential. 

The interaction potential $V(r)$ is parameterized by the fit shown in
Fig.~\ref{fig:interactions}(d) and approximated by the isotropic form
\begin{equation}
V(r) = V_0\,K_0(r/\xi),
\label{eq:potential}
\end{equation}
where $K_0$ is the modified Bessel function of the second kind,
$\xi = 1.159$ is the decay length of the single-skyrmion tail
[Fig.~\ref{fig:single_skyrmion}], and $V_0 = 921$ is fitted to the numerical
data. In what follows, we neglect the magnetic-field dependence of the
interaction potential; the decay length $\xi$, for instance, changes by less
than 2\% between $H = 0.96$ and $H_c$.

Close to the saturation field, $H \lesssim H_c$, the system enters a dilute regime where the skyrmion density is determined by the competition between the {linear energy gain associated with the effective chemical potential,} $\propto (H_c-H)$, and the exponentially weak repulsive interaction in Eq.~\eqref{eq:potential}. This regime is particularly relevant because the typical separation between low-energy pinning centers of type $A$ [see Fig.~\ref{fig:pinning}] is {smaller than} the interaction range $\xi$.

In this limit, the typical skyrmion separation is determined by the balance between the chemical potential and the inter-skyrmion repulsion. Minimizing the total energy then leads to an {essentially singular} equation of state, in which the typical separation grows logarithmically as $H \to H_c^-$. Consequently, the skyrmion density {asymptotically} vanishes as
\begin{equation}
n_{\rm sk} \sim \frac{1}{\left[\ln\!\left(\Lambda/(H_c - H)\right)\right]^2},
\label{Eq:sk_den}
\end{equation}
where $\Lambda$ is a nonuniversal scale set by microscopic parameters. The {diverging derivative} of $n_{\rm sk}(H)$ as $H \to H_c^-$ reflects the exponentially weak nature of the repulsive interaction in the dilute limit.

Numerical minimization of $E_{\mathrm{tot}}$ over the set $\{n_i\}$ provides a tractable method {for approximating} low-energy skyrmion configurations. We use a simulated-annealing approach in the grand-canonical ensemble, as described in the Supplemental Material~\cite{SM}. This approach captures the competition between the local energy gain associated with placing a skyrmion on a negative-energy pinning site and the inter-skyrmion repulsion. The configurations obtained in this way are then used as initial states for a BFGS minimization of the full spin Hamiltonian, Eq.~\eqref{eq:mham}, at the corresponding magnetic field, and the lowest-lying state is selected according to its full spin energy. The approximations of the effective model, such as the point-like treatment of the skyrmions and the neglect of the field dependence of $V(r)$, therefore affect only the selection of candidate configurations, not the final configurations and energies.

\begin{figure}
    \centering
    \includegraphics[width=0.45\textwidth]{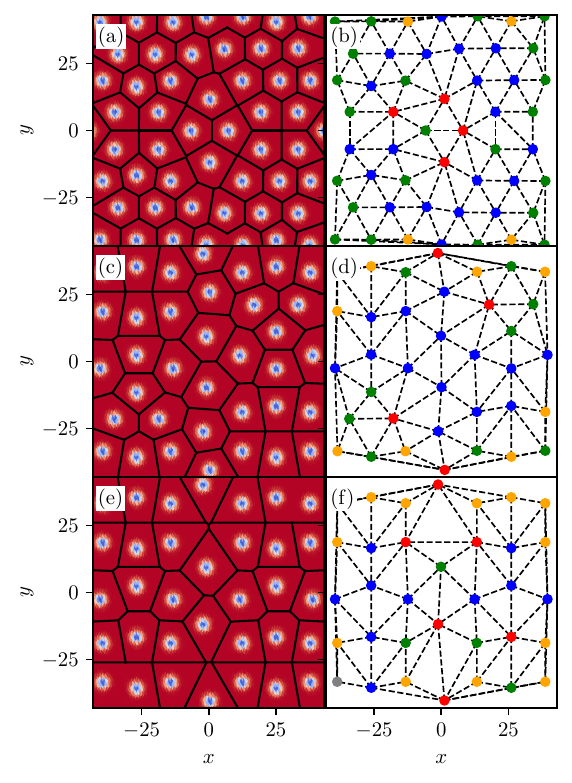}
    \caption{Spin configurations near the critical field, obtained from the effective point-particle model via grand-canonical simulated annealing, on the $100\times100$ quasicrystalline patch ($N=10,700$ sites) at (a,b) $H=0.993$, (c,d) $H=0.994$, and (e,f) $H=0.995$. Left panels show the spin texture with the Voronoi tessellation overlaid; right panels show the corresponding skyrmion network, with sites colored by coordination number.}
    \label{fig:configs}
\end{figure}

Figure~\ref{fig:configs} shows the {low-energy} skyrmion configurations for different values of the magnetic field below the critical field $H_c=0.9957$. The resulting configurations exhibit local triangular and square-lattice correlations, {consistent with those observed in the spin Monte Carlo simulations} [see Fig.~\ref{fig:configs}(a)]. This behavior is expected because the underlying pinning landscape contains {both square and triangular motifs} at different length scales, reflecting the self-similar structure of the quasicrystalline lattice. As the magnetic field approaches $H_c$, the skyrmion density decreases sharply, {consistent with the asymptotic behavior predicted by Eq.~\eqref{Eq:sk_den}.} {At these low densities}, finite-size effects are expected to play a significant role in determining the skyrmion {arrangement} [see Fig.~\ref{fig:configs}(c)].

\begin{figure}[ht]
    \centering
\includegraphics[width=\columnwidth]{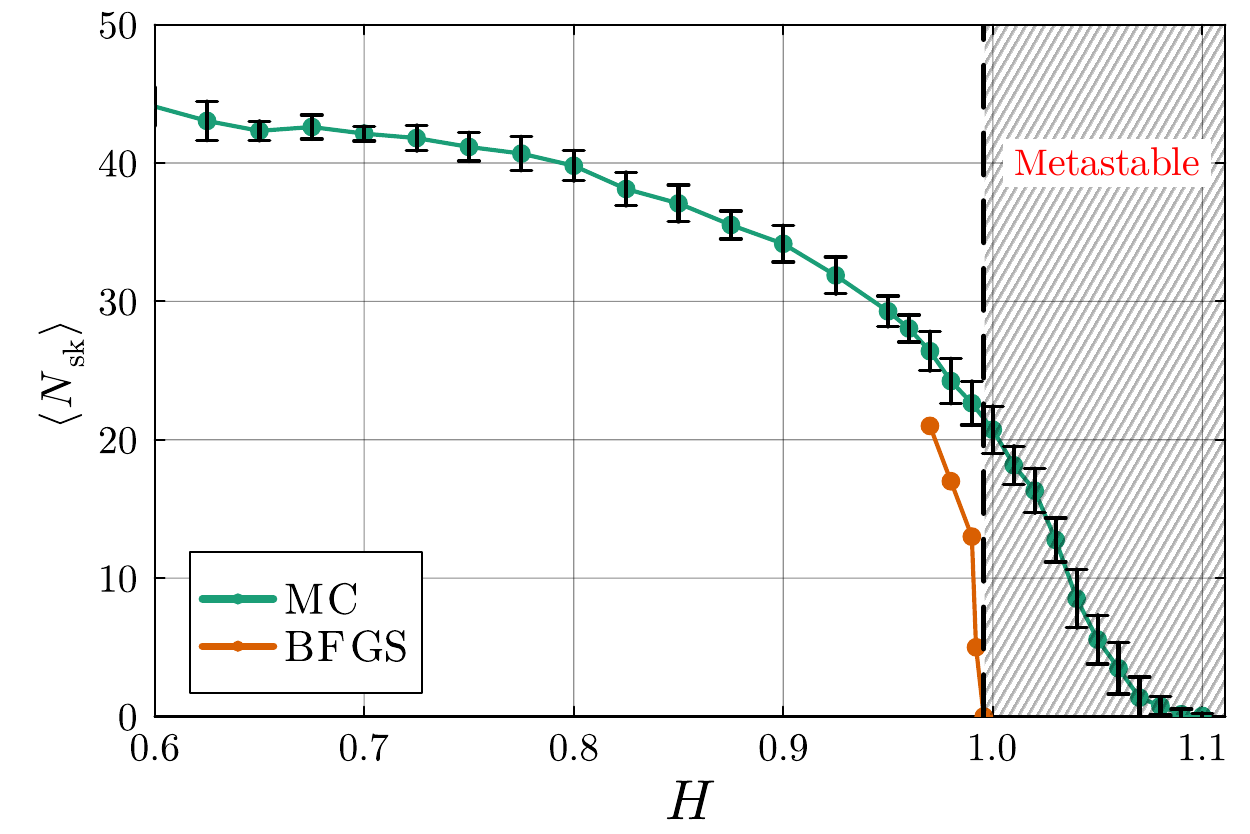}
\caption{Average total skyrmion number ${\langle N_{\rm sk}\rangle=\langle|Q|\rangle}$ as a function of the external field $H$. Green dots: Full spin Monte Carlo simulations (Sec.~\ref{sec:mc}). Orange dots: Effective point-particle model (Sec.~\ref{sec:effmod}), valid for $H\lesssim H_c$. Error bars on the Monte Carlo data denote the standard deviation over 26 independent simulated-annealing replicas at each field. {The shaded region indicates the metastable branch of the Monte Carlo simulations, in which skyrmions persist above $H_c$ because of topological protection.}}
    \label{Fig:Nsk_vsH}
\end{figure}

We have verified that the skyrmion configurations obtained through this method yield {lower energies} than those obtained via spin Monte Carlo simulations for the same set of parameters. This {suggests that the effective particle model captures the essential physics governing low-energy skyrmion configurations, even in the presence of a complex pinning landscape.} Nevertheless, {determining whether the low-density phase develops genuine long-range quasiperiodic order, analogous to the solitonic phases found in one-dimensional chiral quasicrystals}~\cite{cornaglia2024quasicrystalline}, {will require simulations on significantly larger system sizes.} 
Figure~\ref{Fig:Nsk_vsH} shows the average total skyrmion number $\langle N_{\rm sk}\rangle = \langle|Q|\rangle$ as a function of the applied field $H$, obtained from two complementary methods. The green points {represent} the full spin Monte Carlo simulations of Sec.~\ref{sec:mc}, carried out with simulated annealing over the entire field range. The orange points {represent} the effective point-particle model introduced in this section, {which is} valid in the dilute regime $H\lesssim H_c$.

The spin Monte Carlo skyrmion population persists as a metastable phase above $H_c$ (shaded region): once formed, skyrmions are topologically protected and remain trapped behind finite energy barriers, so the annealing protocol cannot {eliminate} them, and $\langle N_{\rm sk}\rangle$ decays only gradually, vanishing around $H\approx 1.1$. By contrast, the effective point-particle model samples low-energy configurations near $H_c$ more efficiently and yields a markedly smaller skyrmion number that drops sharply to zero as $H\to H_c^-$. This sharp decrease is {consistent} with the singular equation of state in Eq.~\eqref{Eq:sk_den}, according to which $n_{\rm sk}$ vanishes {as the inverse square of a logarithm} as $H\to H_c^-$.

\section{Tunable topological Hall conductivity} \label{sec:THE}

In the previous sections we have shown how the quasi-periodicity of the
lattice allows for {continuous tuning of the equilibrium skyrmion density}, as illustrated in
Fig.~\ref{Fig:Nsk_vsH}. 
{While metastable skyrmion configurations persist above $H_c$ because of topological protection, the equilibrium skyrmion density vanishes continuously.} This represents a significant improvement over conventional skyrmion crystals on periodic lattices, where the transition {to the fully polarized state is strongly first order}~\cite{Bogdanov1994,Yu2010,Nagaosa2013,Leonov2015,Batista2016}. The ability to {quasi-continuously tune} the total topological charge has important physical consequences; in particular, it provides a means to control the topological Hall effect (THE)~\cite{Nagaosa2013,Raju2019,Mandal2025}.

To illustrate this possibility, we couple the localized spin texture to an
itinerant electron bath through a local {Hund's} exchange interaction,
\begin{equation}
\begin{split}
    \hat{\mathcal{H}}_{\text{it}} &= -t \sum_{\langle ij\rangle}
    \left[\hat{\mathbf{c}}^{\dagger}_i \sigma_0\,\hat{\mathbf{c}}^{}_{j}
    + \text{h.c.} \right]
    - \frac{J_h S}{2} \sum_i
    \hat{\mathbf{c}}^{\dagger}_i
    \left(\mathbf{n}_i \cdot \vec{\sigma}\right)
    \hat{\mathbf{c}}^{}_i,
    \label{Eq:Kondo_model}
\end{split}
\end{equation}
where $t$ is the hopping amplitude, $J_h$ is the {local Hund's} exchange coupling between the conduction electrons and the classical local moments $S\mathbf{n}_i$, $\vec{\sigma}=(\sigma_1,\sigma_2,\sigma_3)$ is the vector of Pauli matrices, $\sigma_0$ is the $2\times2$ identity matrix, and
\begin{equation}
    \hat{\mathbf{c}}^{}_i =
    \begin{bmatrix}
    \hat{c}^{}_{i\uparrow}\\
    \hat{c}^{}_{i\downarrow}
    \end{bmatrix},
    \qquad
    \hat{\mathbf{c}}^{\dagger}_i =
    \begin{bmatrix}
    \hat{c}^{\dagger}_{i\uparrow} &
    \hat{c}^{\dagger}_{i\downarrow}
    \end{bmatrix}
\end{equation}
are {spinors of} fermionic creation and annihilation operators. We neglect the Zeeman coupling of the itinerant electrons to the external magnetic field, as it is typically much smaller than the electronic bandwidth set by $t$.

{A Hund's coupling $J_h$ comparable to the Fermi energy naturally arises in systems containing both localized and itinerant orbitals on the same atomic site.} In the strong-coupling regime $J_h/t\gg1$, corresponding to the double-exchange limit, the coupling induces an effective ferromagnetic exchange that can be absorbed into the Heisenberg term of the spin Hamiltonian in Eq.~\eqref{eq:mham}~\cite{Blundell2001}. In the following, we neglect the feedback of the itinerant electrons on the magnetic texture and compute the topological Hall effect generated by the magnetic configurations obtained from the pure spin Hamiltonian in Eq.~\eqref{eq:mham}. Results for the {weak-coupling} tight-binding model are presented in the Supplemental Material~\cite{SM}.

{We now evaluate the Hall conductivity associated with the quasicrystalline skyrmion textures.} Since the system is a quasicrystal and therefore lacks translational invariance, the Hamiltonian in Eq.~\eqref{Eq:Kondo_model} must be diagonalized in real space. To eliminate spurious edge modes, we impose periodic boundary conditions on the square-triangle approximant of order $n=3$, {obtained by periodically repeating a square patch of side} $(2+\sqrt{3})^3$ {of the quasicrystal}. {Diagonalization yields} a set of single-particle eigenstates $\ket{n}$ with energies $E_n$. The {Hall} conductivity is evaluated from the Kubo formula for the retarded current-current {correlation function} in the static ($\omega\to0$) limit~\cite{Nittis2017,Mukherjee2023}.

The current operator is defined as $\hat{J}^{\alpha}=q\,\hat{v}^{\alpha}$ with $\alpha=x,y$, where the velocity operator {is obtained} from the lattice continuity equation
\begin{equation}
    \hat{v}^{\alpha}
    = \frac{-\mathrm{i}\,t}{\hbar}
      \sum_{\langle i,j\rangle,\sigma}
      \bigl(r^{\alpha}_{j}-r^{\alpha}_{i}\bigr)
      \Bigl[\hat{c}^{\dagger}_{i\sigma}\hat{c}^{\phantom{\dagger}}_{j\sigma}
      -\hat{c}^{\dagger}_{j\sigma}
       \hat{c}^{\phantom{\dagger}}_{i\sigma}\Bigr]\,.
\end{equation}
The Hall conductivity {is then given by}
\begin{equation}
    \sigma_{xy}
    = \frac{e^{2}\hbar}{A}
      \sum_{n\neq m}
      \frac{f(E_n)-f(E_m)}{(E_n-E_m)^{2}+\eta^{2}}\;
      \operatorname{Im}\!\Bigl[
      \mel{n}{\hat{v}^{x}}{m}\,\mel{m}{\hat{v}^{y}}{n}\Bigr]\,,
    \label{Eq:Kubo}
\end{equation}
where $A$ is the total area of the sample,
$f(E)=\bigl[\mathrm{e}^{(E-\mu)/k_{B}T}+1\bigr]^{-1}$ is the
Fermi-Dirac distribution at temperature~$T$ and chemical
potential~$\mu$, and $\eta$ is a small positive broadening parameter that
{broadens the $\delta$-function peaks of the clean system into Lorentzians of width $\eta$, thereby accounting phenomenologically for a finite quasiparticle lifetime.}
 
\begin{figure}[ht]
    \centering
    \includegraphics[width=\columnwidth]{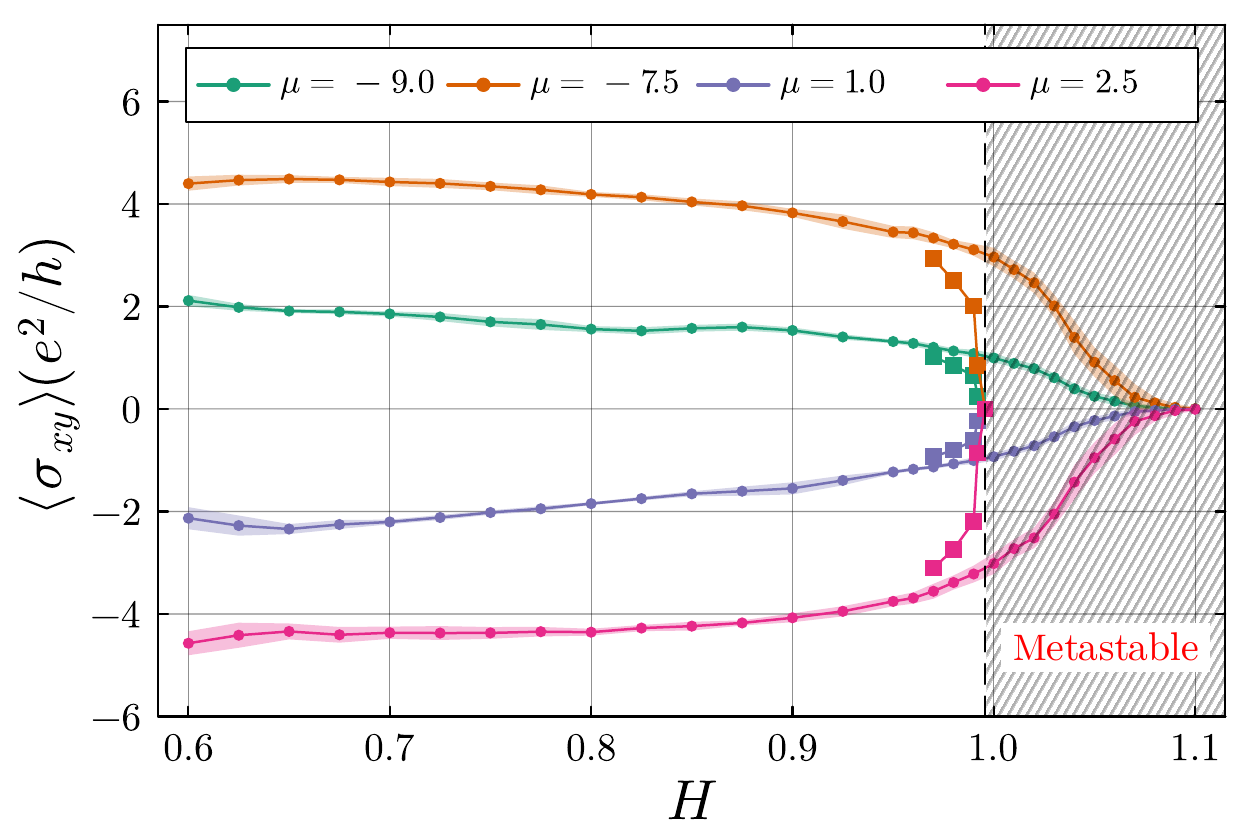}
    \caption{Hall conductivity $\sigma_{xy}$, in units of $e^{2}/h$, {computed from} the MC samples of the periodic approximant as a function of the external field $H$ for several values of the chemical potential $\mu$. Shaded bands indicate the standard deviation {of} the MC samples. Square markers show the same quantity near the saturation field, obtained with the BFGS method. {The shaded region indicates the field range over which the metastable skyrmion phase persists in the Monte Carlo simulations.}}
    \label{Fig:results_sigma}
\end{figure}

We set the hopping amplitude as the unit of energy ($t=1$) and work in the strong-coupling regime with $J_hS=10$. At $T=0$, the Fermi--Dirac distribution reduces to a step function, $f(E)=\Theta(\mu-E)$, and we use a broadening parameter $\eta=0.1\,t$. Exact diagonalization of Eq.~\eqref{Eq:Kondo_model}, followed by the evaluation of the Hall conductivity via Eq.~\eqref{Eq:Kubo} over a range of chemical potentials $\mu$, yields the results shown in Fig.~\ref{Fig:results_sigma}. As anticipated, the {continuous tunability} of the skyrmion number under an applied magnetic field translates into {continuous control} of the Hall conductivity. The effect is most pronounced near the saturation field, where the skyrmion density varies quasi-continuously, allowing the topological Hall response to be tuned smoothly to zero. The dependence of the Hall conductivity on the chemical potential is presented in the Supplemental Material~\cite{SM}.

\section{Summary and conclusions}\label{sec:sumcon}

We have investigated the magnetic properties of a classical Heisenberg model with {DM} interactions on a two-dimensional quasicrystalline lattice composed of squares and equilateral triangles. The quasicrystal structure introduces a quasiperiodic pinning potential that {qualitatively modifies} skyrmion behavior compared to periodic systems.

The system supports metastable single-skyrmion solutions with a well-defined topological charge $|Q|=1$. These skyrmions localize at specific sites determined by the local geometry of the lattice, and their energies form a singular spectrum with multiple nearly degenerate minima. This results in a spatially varying energy landscape with recurrent but non-identical pinning environments.

Monte Carlo simulations reveal different regimes for the skyrmion configurations as a function of magnetic field, {ranging} from triangular lattices dominated by inter-skyrmion repulsion at low fields, to disordered states at intermediate fields, and to sparse, pinned configurations near $H_c$. A statistical analysis of local coordination and bond angles highlights the emergence of both triangular and square-lattice correlations {originating from} the underlying lattice self-similarity.

To capture the low-density regime near $H_c$, we developed an effective model in which skyrmions are treated as point-like particles confined to metastable pinning centers. The model incorporates site-dependent skyrmion energies and isotropic pairwise repulsion, and {reproduces the main features of the low-energy configurations obtained from the full spin simulations. Most importantly, the singular hierarchy of pinning energies generated by the quasiperiodic lattice enables a quasi-continuous suppression of the skyrmion density upon approaching the saturation field, in sharp contrast to the strongly first-order collapse of the skyrmion crystal typically found in periodic lattices.}

Our results show that quasicrystals provide a versatile platform for engineering complex skyrmion configurations through a combination of geometry-induced pinning and inter-skyrmion interactions. This suggests potential for the design of skyrmion-based memory and logic devices with multistable states rooted in the quasiperiodic structure. Moreover, the tunability of the skyrmion density with magnetic field enables {continuous} control of the Hall conductivity of an electron bath coupled to the skyrmion background. In particular, the rapid decrease of the skyrmion density near the saturation field provides an \emph{amplification} mechanism, in which a small change in the applied magnetic field produces a large change in {the Hall response}.

Although our analysis focuses on the low-temperature classical regime, we expect the field tunability of the skyrmion density to persist at finite temperatures where skyrmion textures remain stable. Thermal fluctuations can redistribute skyrmions among nearly degenerate pinning configurations and round the finest features of the field dependence, while preserving the underlying quasiperiodic pinning landscape. Establishing the temperature range over which this tunability persists is an important direction for future work.

Furthermore, skyrmions driven over quasicrystalline substrates are expected to exhibit rich dynamical behavior resulting from the underlying quasiperiodic pinning potential. Previous studies of vortices and colloids interacting with quasicrystalline pinning arrays have shown that the underlying potential can induce directional locking, where particle motion becomes constrained to preferred angles~\cite{reichhardt2011dynamical,reichhardt2012vortex}. Moreover, the system can undergo dynamically induced ordering transitions depending on the drive direction and particle density.

The analogy between vortex systems and the present skyrmion model is further substantiated by the nature of the inter-particle interaction. In studies of driven vortices on quasiperiodic substrates~\cite{reichhardt2011dynamical,reichhardt2012vortex}, the vortex--vortex interaction is modeled by the modified Bessel function of the second kind, $K_1(r/\lambda)$, which decays exponentially for $r\gg\lambda$. In our model, a direct numerical evaluation of the skyrmion--skyrmion interaction confirms that the effective potential $V(r)$ also decays exponentially with separation [see Eq.~(\ref{eq:potential})]. This result is consistent with particle-based models of skyrmions in chiral magnets, where similar exponential interactions are derived under the assumption of rigid skyrmion profiles~\cite{lin2013particle}. {The common exponential form of the interaction potential strongly suggests that skyrmions driven over quasiperiodic substrates should exhibit a similarly rich phenomenology, including directional locking and dynamically induced ordering transitions, as observed in vortices, colloids, and particle-based models of skyrmions~\cite{reichhardt2011dynamical,reichhardt2012vortex,lin2013particle}.}

\begin{acknowledgments}
    L.M.C. was supported by the U.S. Department of Energy, Office of Science, Basic Energy Sciences, Materials Science and Engineering Division. This work was supported by the Consejo Nacional de Investigaciones Científicas y Técnicas (CONICET), Argentina, under PIP Grant No. 11220250100515CO. We acknowledge the Texas Advanced Computing Center (TACC) for providing the computational hardware and the Physics Department at Centro Atómico Bariloche (CNEA) for hosting and operating the HPC cluster used in this work. F. A. G. A. is partially supported by SECyT UNLP PI+D X947 and X1065, and Grant PICT-2020 - SERIEA-03205 of the ANPCyT. This collaborative effort is made possible from generous support by Instituto Balseiro, Universidad de Cuyo, Fulbright Argentina, and through a Fulbright Specialist award. The Fulbright Specialist Program, a part of the larger Fulbright Program, is a program of the Department of State Bureau of Educational and Cultural Affairs with funding provided by the U.S. Government and administered by World Learning. C. D. B. also acknowledges support from the U.S. Department of Energy, Office of Science, Office of Basic Energy Sciences, under Award Number DE-SC0022311.
\end{acknowledgments}

\bibliography{2DQC}
\end{document}


\title{Supplemental Material: Tunable Emergent Gauge Fields from Skyrmions in a Quasicrystalline Lattice}

\author{Leandro M. Chinellato}
\email[]{lchinell@vols.utk.edu}
\affiliation{Department of Physics and Astronomy, The University of Tennessee, Knoxville, TN, 37996, USA}
\author{Flavia A.  G\'omez Albarrac\'in}
\email[]{albarrac@fisica.unlp.edu.ar}
\affiliation{Instituto de F\'isica de L\'iquidos y Sistemas Biol\'ogicos, CONICET, Facultad de Ciencias Exactas, Universidad Nacional de La Plata, 1900 La Plata, Argentina}
\affiliation{Departamento de Ciencias B\'asicas, Facultad de Ingenier\'ia, UNLP, La Plata, Argentina}
\author{Cristian D. Batista}
\affiliation{Department of Physics and Astronomy, The University of Tennessee, Knoxville, TN, 37996, USA}
\affiliation{Neutron Scattering Division and Shull-Wollan Center, Oak Ridge National Laboratory, Oak Ridge, TN, 37831, USA}
\author{Pablo S. Cornaglia~\orcidlink{0000-0003-4991-6573} }
\email[]{pablo.cornaglia@ib.edu.ar}
\affiliation{Centro Atómico Bariloche and Instituto Balseiro, CNEA, 8400 Bariloche, Argentina}
\affiliation{Consejo Nacional de Investigaciones Científicas y Técnicas (CONICET), Argentina}
\affiliation{Instituto de Nanociencia y Nanotecnología CNEA-CONICET}

\maketitle

\section{\label{Sec:unfolding}Unfolding of the dynamical spin structure factor}

As shown in the main text, the Dzyaloshinskii--Moriya
interaction does not contribute to the spin-wave Hamiltonian at
quadratic order around the fully polarized state. {The resulting
single-magnon Hamiltonian} is therefore given by
\begin{equation}\label{eq:dynmat}
  (\mathcal{H}_{\rm sw})_{ij} =
  \begin{cases}
    z_i SJ + H & i = j, \\
    -SJ         & j \in \mathrm{nn}(i), \\
    0          & \text{otherwise}
  \end{cases}
\end{equation}
where $z_i$ is the coordination number of site $i$ and $\mathrm{nn}(i)$ denotes its set of nearest neighbors. Equation~(\ref{eq:dynmat}) {is mathematically equivalent to} a single-particle tight-binding Hamiltonian on the quasicrystalline lattice, with a site-dependent on-site energy $\mu_i=z_iSJ+H$ arising from the non-uniform local coordination environment inherent to quasicrystalline tilings, {allowing us to directly exploit the well-established phenomenology of localization and criticality in quasiperiodic systems}. We diagonalize $\mathcal{H}_{\rm sw}$ numerically for a commensurate approximant of size $N=2911$ and obtain the complete set of eigenstates $\{|\psi_n\rangle\}_{n=1}^{N}$ with eigenvalues~$\omega_n$.

\begin{figure}[H]
    \centering
    \includegraphics[width=\linewidth]{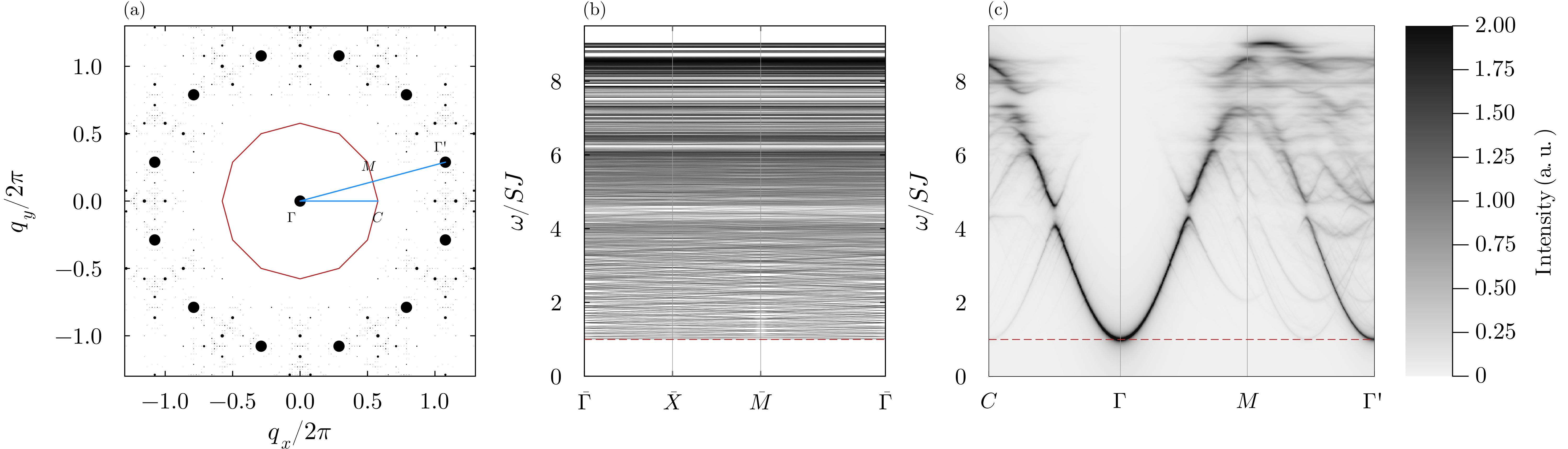}
    \caption{(a) Lattice structure factor $I(\mathbf{q})$ of the $N=2911$ approximant; circles mark the first-order peaks $\mathbf{g}^{I}$ and the dodecagon is the pseudo-Brillouin zone. (b) Folded one-magnon spectrum $\omega_n(\mathbf{K})$ in the supercell Brillouin zone. (c) Unfolded one-magnon dynamical structure factor $\mathcal{S}^{\perp}(\mathbf{q},\omega)$ along $C$--$\Gamma$--$M$--$\Gamma'$ of the pseudo-Brillouin zone ($M=\mathbf{g}^{I}/2$, $\Gamma'=\mathbf{g}^{I}$), for $H=0.9957J$ and $\eta=0.03J S$; the dashed line marks $\omega=H$.}
    \label{fig:sqw}
\end{figure}

The approximant is a square supercell of side $L=(2+\sqrt{3})^3$ repeated periodically, so Eq.~(\ref{eq:dynmat}) is the $\mathbf{K}=0$ block of the Bloch Hamiltonian
\begin{equation}\label{eq:bloch}
  \mathcal{H}_{\rm sw}(\mathbf{K})_{ij}=(z_iSJ+H)\,\delta_{ij}-JS\sum_{\mathbf{d}_{ij}}e^{i\mathbf{K}\cdot\mathbf{d}_{ij}},
\end{equation}
where $\mathbf{d}_{ij}$ are the actual bond vectors, including those that cross the supercell boundary, and $\mathbf{K}$ lies in the supercell Brillouin zone, of linear size $2\pi/L\simeq0.12$. Its eigenvalues $\omega_n(\mathbf{K})$ form $N=2911$ bands folded into this small zone [Fig.~\ref{fig:sqw}(b)]: the folded spectrum is dense and nearly flat, and carries no information on the momentum $\mathbf{q}$ probed in a scattering experiment. We recover this information by band unfolding, as presented in Ref.~\cite{Zhang2022}.

The idea is to measure how much plane-wave character $e^{i\mathbf{q}\cdot\mathbf{r}}$ each supercell eigenstate carries. The eigenvectors $u_n(j;\mathbf{K})$ of Eq.~(\ref{eq:bloch}) are the cell-periodic parts of the Bloch eigenstates, whose amplitude on site $j$ of the supercell at lattice vector $\mathbf{R}$ is $e^{i\mathbf{K}\cdot(\mathbf{r}_j+\mathbf{R})}u_n(j;\mathbf{K})$. Writing $\mathbf{q}=\mathbf{K}+\mathbf{G}$, with $\mathbf{G}$ a supercell reciprocal lattice vector, and using $e^{i\mathbf{G}\cdot\mathbf{R}}=1$, the plane wave $|\mathbf{q}\rangle$ of the main text lies entirely in the $\mathbf{K}$ sector, where it reduces to $\phi_{\mathbf{G}}(j)=N^{-1/2}e^{i\mathbf{G}\cdot\mathbf{r}_j}$, with $j$ running over one supercell. The matrix element in Eq.~(5) of the main text therefore becomes
\begin{equation}\label{eq:resolvent}
\langle\mathbf{q}|\big(\omega+i0^{+}-\mathcal{H}_{\rm sw}\big)^{-1}|\mathbf{q}\rangle
=\langle\phi_{\mathbf{G}}|\big(\omega+i0^{+}-\mathcal{H}_{\rm sw}(\mathbf{K})\big)^{-1}|\phi_{\mathbf{G}}\rangle .
\end{equation}
Expanding in the eigenvectors $u_n(\mathbf{K})$ gives
\begin{equation}\label{eq:dssf}
  \mathcal{S}^{\perp}(\mathbf{q},\omega)=S\sum_n P_n(\mathbf{q})\,\delta\big(\omega-\omega_n(\mathbf{K})\big),
\end{equation}
where
\begin{equation}\label{eq:weight}
  P_n(\mathbf{q})=\big|\langle\phi_{\mathbf{G}}|u_n(\mathbf{K})\rangle\big|^2=\Big|\frac{1}{\sqrt{N}}\sum_j e^{-i\mathbf{G}\cdot\mathbf{r}_j}\,u_n(j;\mathbf{K})\Big|^2
\end{equation}
is the spectral weight of band $n$ at $\mathbf{q}$. It satisfies $\sum_n P_n(\mathbf{q})=1$, so that $\int\mathcal{S}^{\perp}d\omega=S$. For a crystal with one site per primitive cell, $P_n(\mathbf{q})$ is either $0$ or $1$ and the procedure reproduces the primitive band structure exactly. In the approximant the weight of a given $\mathbf{q}$ is shared among several folded bands, and plotting each band $\omega_n(\mathbf{K})$ at $\mathbf{q}=\mathbf{K}+\mathbf{G}$ with weight $P_n(\mathbf{q})$ yields the unfolded spectrum. Rather than diagonalizing Eq.~(\ref{eq:bloch}) at every $\mathbf{q}$, we evaluate Eq.~(\ref{eq:resolvent}) with a Lanczos continued fraction~\cite{Gagliano1987} started from $|\phi_{\mathbf{G}}\rangle$ (600 steps, Lorentzian broadening $\eta=0.03JS$).

A quasicrystal has no Brillouin zone, but its diffraction pattern defines a pseudo-Brillouin zone~\cite{Zhang2022}. It is bounded by the bisectors of the twelve strongest peaks $\mathbf{g}^{I}$ of the lattice structure factor $I(\mathbf{q})=N^{-2}\big|\sum_j e^{-i\mathbf{q}\cdot\mathbf{r}_j}\big|^2$, located at $|\mathbf{g}^{I}|/2\pi=1.115$ and angles $15^\circ+30^\circ m$ [Fig.~\ref{fig:sqw}(a)]. We plot $\mathcal{S}^{\perp}(\mathbf{q},\omega)$ along $C$--$\Gamma$--$M$--$\Gamma'$, where $M=\mathbf{g}^{I}/2$ is an edge midpoint, $C$ a corner, and $\Gamma'=\mathbf{g}^{I}$ lies in the neighboring zone.
 
Figure~\ref{fig:sqw}(c) shows that, in contrast to the folded spectrum, the spectral weight is concentrated in a sharp, dispersive branch starting at $\omega=H$ at $\Gamma$, where the uniform mode is an exact eigenstate, with $\omega\simeq H+JS\bar{z}|\mathbf{q}|^2/4$ and average coordination $\bar{z}=5.07$. The branch shows an avoided crossing at $4.1JS\lesssim\omega\lesssim4.7JS$, where it meets its replica shifted by $\mathbf{g}^{I}$, and the replica returns to $\omega=H$ at $\Gamma'=\mathbf{g}^{I}$ with weight $0.306\,S$, equal to $I(\mathbf{g}^{I})$. Above $\omega\approx7.5JS$ the weight spreads over faint, weakly dispersive features, consistent with the multifractal high-energy eigenstates discussed in Sec.~\ref{Sec:IPR}.

\section{\label{Sec:IPR}Inverse participation ratio analysis of the magnon eigenstates}

To quantify the spatial extent of the magnon eigenstates on the
quasicrystalline lattice, we compute the inverse participation ratio
(IPR) of the eigenstates $|\psi_n\rangle$ of Eq.~(\ref{eq:dynmat}).

For each normalized eigenstate, the IPR is defined as~\cite{BKramer1993}
\begin{equation}\label{eq:ipr}
  \mathrm{IPR}_n = \sum_{i=1}^{N} |\psi_n(i)|^4\,,
\end{equation}
where $\psi_n(i)=\langle i|\psi_n\rangle$ is the amplitude on
site~$i$. {The IPR provides a quantitative measure of the spatial extent of an eigenstate.}
A state uniformly extended over all~$N$ sites satisfies
$|\psi_n(i)|^2=1/N$, giving
$\mathrm{IPR}_n=1/N\to0$ in the thermodynamic limit.
Conversely, a state strictly localized on $\ell$ sites has
$\mathrm{IPR}_n\sim1/\ell$, which remains of order unity as
$N\to\infty$. {Intermediate power-law scaling,}
$\mathrm{IPR}_n\sim N^{-\alpha}$ with $0<\alpha<1$,
{is characteristic of} multifractal (critical) eigenstates.

\begin{figure}[H]
    \centering
    \includegraphics[width=0.5 \linewidth]{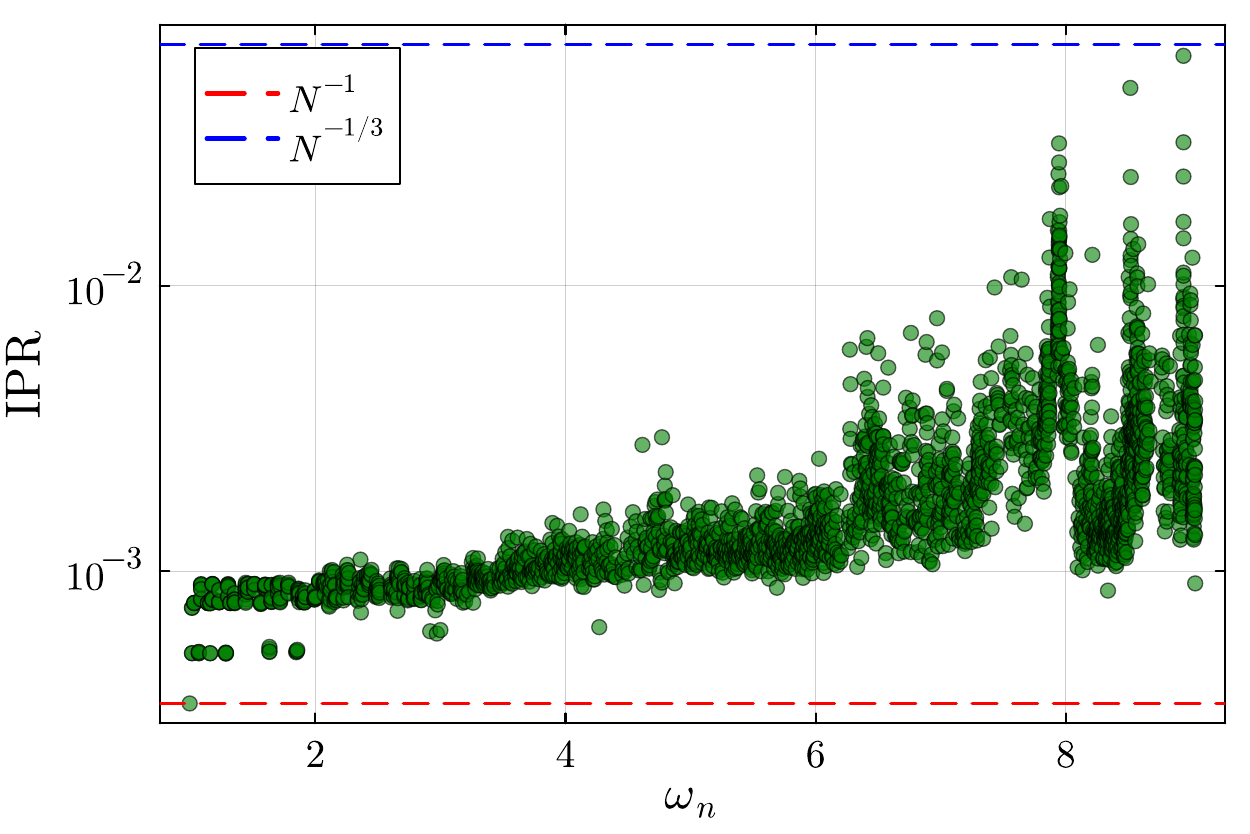}
    \caption{Inverse participation ratio (IPR) of the tight-binding eigenstates for an approximant with $N=2911$ sites, plotted as a function of the eigenenergy~$\omega_n$ (in units of~$JS$). The dashed horizontal lines {mark} the extended-state value, $\mathrm{IPR}=N^{-1}$ (red), and the {upper bound of the critical regime}, $\mathrm{IPR}=N^{-1/3}$ (blue).}
    \label{fig:IPR}
\end{figure}

Fig.~\ref{fig:IPR} shows the IPR as a function of the eigenenergy~$\omega_n$.
The {lowest-energy magnon state} has $\mathrm{IPR}\sim1/N$, confirming that
the corresponding eigenstate is fully extended across the lattice.
The higher-energy states exhibit {intermediate power-law scaling},
$\mathrm{IPR}\sim N^{-\alpha}$ with $\alpha\in[\tfrac{1}{3},1]$,
signaling multifractal eigenstates that are neither fully extended nor
fully localized. Such critical behavior is generically expected for
tight-binding models on quasicrystalline tilings~\cite{Suck2013}
{and is therefore consistent with the critical nature of magnon eigenstates on the underlying quasiperiodic lattice.}

\color{black}

\section{Skyrmion tail and skyrmion--skyrmion interaction}
\label{sec:linear-response}

Far from its core, a skyrmion is a small perturbation of the quasi-fully polarized (QFP) state. Its tail and the interaction between well-separated skyrmions can therefore be described by linear response. We use the continuum results of Ref.~~\cite{BartonSinger2023}, with effective parameters appropriate to our lattice.

We measure the perturbation relative to the local QFP direction. We define
\begin{equation}
  \theta_i=\arccos\!\left(
  \frac{\mathbf S_i\cdot\mathbf S_i^{\rm QFP}}{S^2}\right).
  \label{eq:lr-angle}
\end{equation}
which removes the weak, spatially varying canting already present without a skyrmion. 

To leading order in this background canting, the transverse response is governed by the spin-wave matrix of Eq.~(\ref{eq:dynmat}). This approximation neglects the corrections to the quadratic energy generated by expanding about the canted QFP state. 
 
We describe the tail by the isotropic quadratic energy
\begin{equation}
  E_{\rm lin}=\frac{\rho S c_{\rm eff}}{2}
  \int d^2r\,\left[|\nabla\psi|^2+\xi^{-2}|\psi|^2\right],
  \label{eq:lr-map}
\end{equation}
where $\psi=(\delta S_x+i\delta S_y)/S$ is the transverse deviation from the QFP background,
$c_{\rm eff}=JS\bar z/4$, $\xi\simeq\sqrt{c_{\rm eff}/H}$,  and $\rho=(1+2/\sqrt3)/2\simeq1.077$ is the site density.
Equation~\eqref{eq:lr-map} is the zero-gauge-field case of the linearized functional in Ref.~\cite{BartonSinger2023}. This identification uses the nearly polarized background and the in-plane DM vectors.

Outside the core, $(-\nabla^2+\xi^{-2})\psi=0$.
For an axisymmetric skyrmion, the leading isotropic tail is (see Ref.~\cite{BartonSinger2023})
\begin{equation}
  \theta(r)\simeq C K_1(r/\xi)
  \label{eq:lr-tail}
\end{equation}
Here $K_\nu$ denotes the modified Bessel function of the second kind.

For two identical axisymmetric skyrmions with the same helicity, the asymptotic interaction follows directly from Ref.~\cite{BartonSinger2023}:
\begin{equation}
  V(R)\simeq V_0K_0(R/\xi),\qquad
  V_0=2\pi\rho S c_{\rm eff}C^2.
  \label{eq:lr-V}
\end{equation}
It is repulsive and has the same exponential decay length as the transverse tail:
$V(R)\simeq V_0\sqrt{\pi\xi/(2R)}e^{-R/\xi}$.
The quasicrystalline environment can modify the tail amplitude and introduce angular dependence. Accordingly, we use Eq.~\eqref{eq:lr-V} as an isotropic approximation.

\section{Single-skyrmion profiles: dependence on field, $D/J$ and system size}\label{sm:profiles}

\subsection{Method}

We computed isolated skyrmions for $J=1$ and $D/J=0.5$, $0.75$, $1.08$ and $1.5$, at different values of the field $H$ that span the full range in which a single skyrmion is metastable, from below the critical field $H_c$ (where $\varepsilon_s=0$) up to its collapse. 
The main calculations use the periodic approximant of the main text: the level-3 inflation with $2911$ sites, and periodic boundary conditions (PBC).
For each $D$ the skyrmion is placed at three pinning centers near the center of the cell. These belong to the three lowest-energy groups A, B and C of the pinning-energy distribution of Sec.~IV (main text).

\textit{Nucleation and field continuation.}
Each skyrmion is nucleated by reversing the spins within a radius $1.6\,J/D$ of the pinning center in the relaxed quasi-fully polarized (QFP) state, and then relaxing.
The metastable branch is then followed in both directions in $H$. Each field starts from the relaxed skyrmion at the previous field, so the skyrmion stays in the same local minimum until it collapses.

\textit{Minimization.}
The energy is minimized with BFGS, as in the main text.
BFGS stops when the relative energy change per iteration falls below $10^{-15}$. This leaves residual gradients of order $10^{-5}$ along the soft translational mode of the pinned skyrmion, where energy changes are below double-precision resolution for $E\sim10^4J$.
Every state is therefore refined with the fast inertial relaxation engine (FIRE)~\cite{bitzek2006}, which uses only the torques $\mathbf{S}_i\times\partial\mathcal{H}/\partial\mathbf{S}_i$, until the largest torque is below $10^{-11}J$.
This refinement changes the energies by $\sim 10^{-10}J$. 

\textit{Profile analysis.}
As in the main text, $\theta_i$ is the angle between $\mathbf{S}_i$ and the QFP spin at the same site and field.
The core is described by the domain-wall profile~\cite{romming2015}
\begin{equation}\label{sm:eq:dw}
\theta(r)=2\arctan\!\left[\frac{\sinh(R/w)}{\sinh(r/w)}\right].
\end{equation}
 We fit Eq.~\eqref{sm:eq:dw} to all sites with $r<8.6\,J/D$, with the radius $R$, the wall width $w$ and the center position all free. The fitted center defines $r$ in all panels.
The tail is fitted to $\theta=C\,K_1(r/\xi)$ with both $C$ and $\xi$ free. The fit window runs from $r_{\min}=2.7\,J/D$ to the first shell where $\theta$ reaches $10^{-6}$, and does not extend beyond $0.45\,L$.

\textit{Boundary conditions and system size.}
To check finite-size and boundary effects, the same calculations were repeated near $H_c$ for each $D$ on the level-4 $100\times100$ patch ($10\,700$ sites). We used open boundaries and, for $D=1.08$, also fixed boundary spins (the sites within one lattice spacing of the edge held along $\hat z$).
The center of this patch is the periodic approximant rotated by $90^\circ$, so the same three pinning centers are available.

\subsection{Results}

\begin{figure}[H]
\centering
\includegraphics[width=\textwidth]{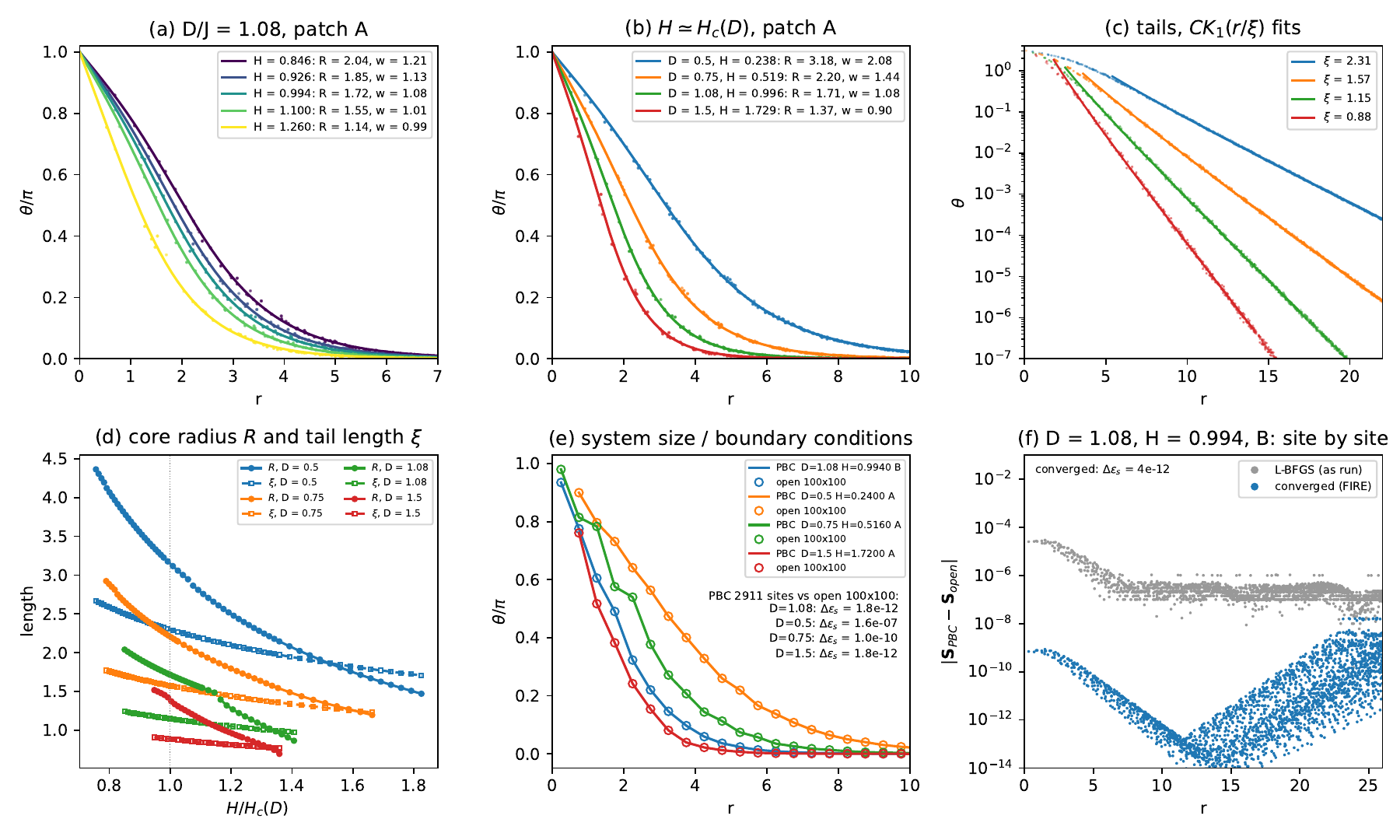}
\caption{Single-skyrmion profiles on the periodic approximant.
(a) $\theta(r)$ at $D/J=1.08$ for several fields (pinning center A). Dots are individual sites and lines are fits to Eq.~\eqref{sm:eq:dw}.
(b) Same for the four values of $D/J$, each at the grid field closest to $H_c(D)$.
(c) Data of (b) on a logarithmic scale, with fits to $C\,K_1(r/\xi)$ ($C$ and $\xi$ free).
(d) Core radius $R$ (circles) and tail length $\xi$ (squares) along the metastable branch versus $H/H_c(D)$.
(e) Shell-averaged $\theta(r)$ for the periodic approximant (lines) and for the open $100\times100$ patch (circles), with the difference in $\varepsilon_s$ between the two.
(f) Site-by-site difference between the two systems at $D/J=1.08$, $H=0.994$ (center B), with distances from the skyrmion center in the periodic cell. Grey: BFGS. Blue: after FIRE refinement.}
\label{sm:fig:profiles}
\end{figure}

Figure~\ref{sm:fig:profiles} summarize the profiles. 
Equation~\eqref{sm:eq:dw} describes the core well for all $D$ and $H$. The scatter of individual sites around it reflects the local environment [Fig.~\ref{sm:fig:profiles}(a,b)].
At fixed $D$, the core radius decreases with $H$, while the wall width changes much less.
The tail length $\xi$ also decreases with $H$ [Fig.~\ref{sm:fig:profiles}(d)] and it agrees with $\xi_{\rm th}=\sqrt{JS\bar z/4H}$, which follows from the magnon spectrum, to within a few percent.

At $H\simeq H_c(D)$, $R$ and $w$ decrease with $D$ roughly as $J/D$, as for skyrmions in continuum chiral magnets. The deviations grow as the skyrmion approaches the lattice scale: 
Since $H_c$ grows with $D$, the tail length also decreases with $D$ [Fig.~\ref{sm:fig:profiles}(c)].

On the $100\times100$ patch the skyrmion profiles coincide with those on the periodic approximant [Fig.~\ref{sm:fig:profiles}(e)].
The difference in $\varepsilon_s$ is below $1.6\times10^{-7}J$. At $D=1.08$ the fixed-spin and open patches give the same $\varepsilon_s$ as the PBC cell to $10^{-7}J$.
Site by site, the remaining difference at $D=1.08$ is about $10^{-9}$. 
The periodic approximant is therefore large enough for all four values of $D/J$. The largest correction occurs for the largest skyrmion ($D=0.5$, $L/2\xi\approx11$), and it is still negligible.

\color{black}
\section{Simulated Annealing Algorithm for Skyrmion Configuration Optimization} \label{app:minen}

We employ a simulated annealing approach to minimize the total energy functional
\begin{equation}
E_{\mathrm{tot}} = \sum_i n_i\, \varepsilon_s
(\mathbf{R}_i; H) + \frac{1}{2} \sum_{i \neq j} n_i n_j\, V\left( |\mathbf{R}_i - \mathbf{R}_j| \right),
\end{equation}
where $ n_i \in \{0,1\} $ indicates the presence of a skyrmion at site $ i $, $ \varepsilon_s(\mathbf{R}_i; H) $ is the effective site energy at field $ H $, and $ V(r) = V_0 K_0(r/\xi) $ is the inter-skyrmion repulsive potential.
 
\paragraph{Input:}
\begin{itemize}
    \item Pinning sites $ \{ \mathbf{R}_i, \varepsilon_s(\mathbf{R}_i;H_c) \} $
    \item External field $ H $ and shift parameter $ \alpha $
    \item Annealing parameters: initial temperature $ T_0 $, cooling rate $ \gamma $, outer steps $ N_s $, inner steps $ N_t $
    \item Interaction parameters: $ V_0 $, $ \xi $
\end{itemize}

\paragraph{Output:}
An approximate configuration $ \{n_i\} $ minimizing $ E_{\mathrm{tot}} $.

\paragraph{Algorithm:}

\begin{enumerate}
    \item \textbf{Filter Sites:} Discard any site $ i $ with $ \varepsilon_s(\mathbf{R}_i; H) = \varepsilon_s(\mathbf{R}_i,H_c) + \alpha (H - H_c) > 0 $.
    
    \item \textbf{Initialize:}
    \begin{itemize}
        \item Set initial configuration $ \mathcal{C}_{\text{current}} \leftarrow \emptyset $
        \item Set best configuration $ \mathcal{C}_{\text{best}} \leftarrow \mathcal{C}_{\text{current}} $
        \item Set best energy $ E_{\text{best}} \leftarrow \infty $
        \item Set temperature $ T \leftarrow T_0 $
    \end{itemize}
     
    \item \textbf{Annealing Loop:} For each outer step $ s = 1, \dots, N_s $
    \begin{itemize}
        \item For each inner step $ t = 1, \dots, N_t $
        \begin{enumerate}
            \item Propose a new configuration $ \mathcal{C}' $ from $ \mathcal{C}_{\text{current}} $ by:
            \begin{itemize}
                \item Adding a skyrmion at an unoccupied site, or
                \item Removing a skyrmion, or
                \item Swapping a skyrmion to another unoccupied site
            \end{itemize}
            \item Compute total energy $ E(\mathcal{C}') $
            \item Accept $ \mathcal{C}' $ with probability
            \[
            p = \min\left(1, \exp\left( -\frac{E(\mathcal{C}') - E(\mathcal{C}_{\text{current}})}{T} \right) \right)
            \]
            \item If accepted, update $ \mathcal{C}_{\text{current}} \leftarrow \mathcal{C}' $
            \item If $ E(\mathcal{C}') < E_{\text{best}} $, update
            \[
            \mathcal{C}_{\text{best}} \leftarrow \mathcal{C}', \quad E_{\text{best}} \leftarrow E(\mathcal{C}')
            \]
        \end{enumerate}
        \item Update temperature: $ T \leftarrow \gamma T $
    \end{itemize}
    
    \item \textbf{Return:} Final configuration $ \mathcal{C}_{\text{best}} $ and its energy $ E_{\text{best}} $.
\end{enumerate}

\section{Tight-binding model}

\begin{figure}[H]
    \centering
    \includegraphics[width=0.5\linewidth]{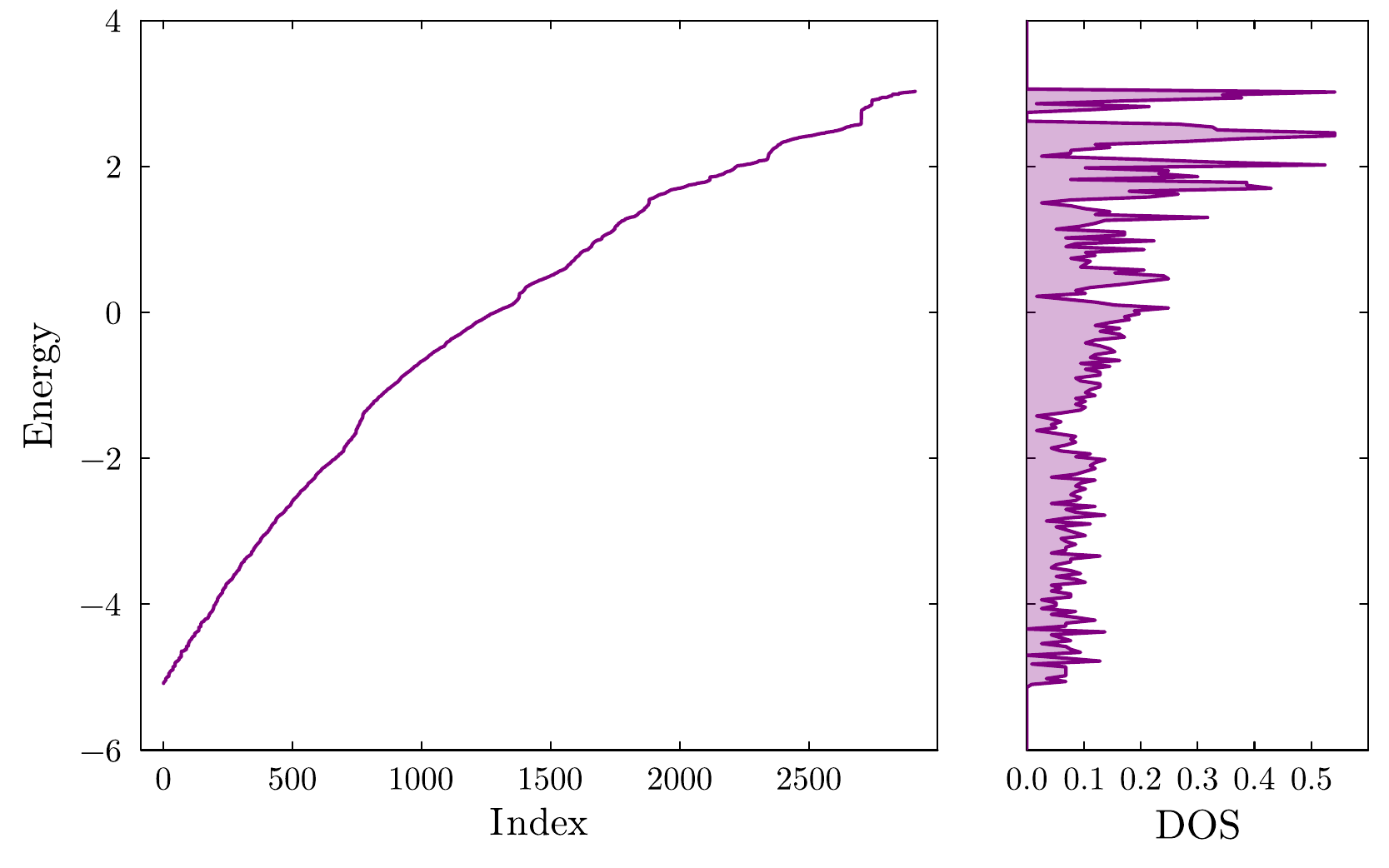}
    \caption{{Eigenvalue spectrum} (left) and corresponding density of states (DOS, right) for the pure tight-binding model on the quasicrystalline lattice. {The DOS is displayed sideways so that both panels share the same energy axis.}}
    \label{Fig:TB_model}
\end{figure}

In this appendix we analyze the electronic structure of the tight-binding model on the quasicrystalline lattice. Since both spin projections are degenerate in this limit, it suffices to diagonalize a single $N\times N$ Hamiltonian for one spin flavor. {The only difference with the one-magnon Hamiltonian discussed in Sec.~\ref{Sec:IPR} is the absence of the site-dependent diagonal term, so both problems share the same hopping matrix and therefore reduce to the same single-particle eigenvalue problem.} Figure~\ref{Fig:TB_model} shows the resulting eigenvalue spectrum (left) together with the corresponding density of states (DOS, right), the latter displayed sideways so that energy is shared as the common vertical axis. {The spectrum and DOS are} in good agreement with Ref.~\cite{Koga2024}.

\section{Hall conductivity curves}

\begin{figure}[H]
    \centering
    \includegraphics[width=0.5\linewidth]{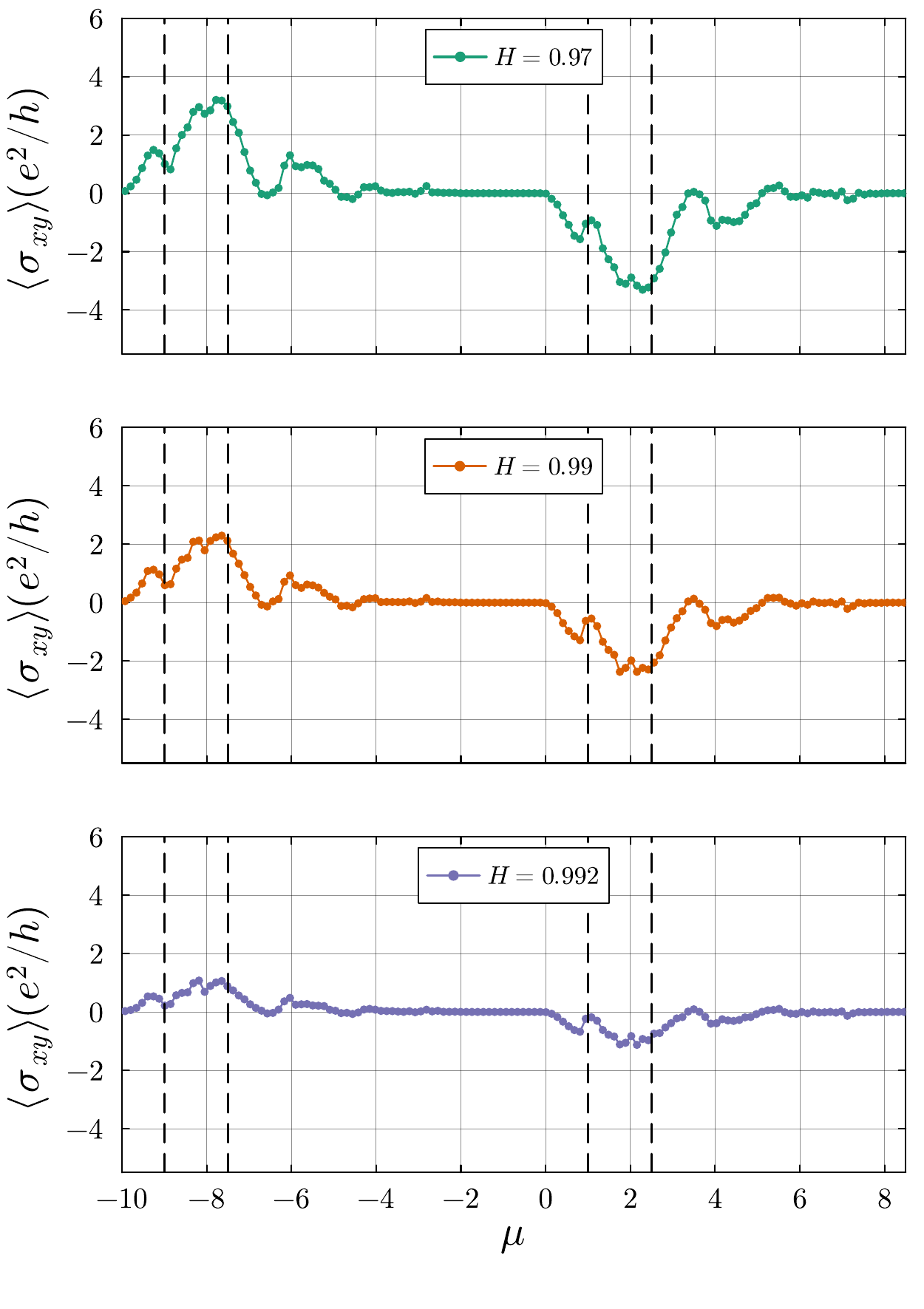}
    \caption{Hall conductivity $\sigma_{xy}$ as a function of the chemical potential~$\mu$ for selected values of the external magnetic field~$H$. The dashed vertical lines {indicate} the values of $\mu$ presented in the main text.}
    \label{fig:sigma_curves}
\end{figure}

In this section we present the Hall conductivity $\sigma_{xy}(\mu,H)$ as a function of the chemical potential~$\mu$ for several values of the external magnetic field~$H$, shown in Fig.~\ref{fig:sigma_curves}. {The Hall conductivity evolves smoothly with magnetic field and is progressively suppressed as the saturation field is approached, reflecting the quasi-continuous decrease of the skyrmion density discussed in the main text.}

\bibliography{2DQC}